%% file: WQSP-IEEE-double-format-arXiv.tex
\documentclass[journal]{IEEEtran}
\input{preambleT.tex}
\usepackage[utf8]{inputenc}
\usepackage[T1]{fontenc}
\usepackage{lmodern}
\usepackage[figurename=Fig.,labelfont=bf,labelsep=period]{caption}

\makeatletter
\def\endthebibliography{%
	\def\@noitemerr{\@latex@warning{Empty `thebibliography' environment}}%
	\endlist
}
\makeatother

\begin{document}
	\title{\LARGE  Revisiting the Water Quality Sensor Placement Problem:\\ Optimizing Network Observability and State Estimation Metrics}
	\author{Ahmad F. Tah$\text{a}^{\dagger,c}$, Shen Wan$\text{g}^\dagger$, Yi Gu$\text{o}^\ddagger$, Tyler H. Summer$\text{s}^\ddagger$, Nikolaos Gatsi$\text{s}^\dagger$, Marcio H. Giacomon$\text{i}^\mathsection$, and  Ahmed A. Abokifa$^*$ \thanks{
			$^c$Corresponding author. $^\dagger$Department of Electrical and Computer Engineering, The University of Texas at San Antonio, TX 78249.   $^{\ddagger}$Department of Mechanical Engineering, The University of Texas at Dallas, Richardson, TX 75080, USA. $^\mathsection$Department of Civil and Environmental Engineering, The University of Texas at San Antonio, TX 78249. $^*$Department of Civil, Materials, and Environmental Engineering, The University of Illinois at Chicago. Emails: ahmad.taha@utsa.edu, mvy292@my.utsa.edu, yi.guo2@utdallas.edu, nikolaos.gatsis@utsa.edu, marcio.giacomoni@utsa.edu,tyler.summers@utdallas.edu, abokifa@uic.edu. This material is based upon work supported by the National Science Foundation under Grants 1728629, 1728605, 2015671, and 2015603.  }} 
	\maketitle
	\begin{abstract}
		Real-time water quality (WQ) sensors in water distribution networks (WDN) have the potential to enable network-wide observability of water quality indicators, contamination event detection, and closed-loop feedback control of WQ dynamics. To that end, prior research has investigated a wide range of methods that guide the geographic placement of WQ sensors. These methods assign a metric for fixed sensor placement (SP) followed by \textit{metric-optimization} to obtain optimal SP. These metrics include minimizing intrusion detection time, minimizing the expected population and amount of contaminated water affected by an intrusion event.  In contrast to the literature, the objective of this paper is to provide a computational method that considers the overlooked metric of state estimation and network-wide observability of the WQ dynamics. This metric finds the optimal WQ sensor placement that minimizes the state estimation error via the Kalman filter for noisy WQ dynamics---a metric that quantifies WDN observability. To that end, the state-space dynamics of WQ states for an entire WDN are given and the observability-driven sensor placement algorithm is presented. The algorithm  takes into account the time-varying nature of WQ dynamics due to changes in the hydraulic profile---a collection of hydraulic states including heads (pressures) at nodes and flow rates in links which are caused by a demand profile over a certain period of time. Thorough case studies are given, highlighting key findings, observations, and recommendations for WDN operators. Github codes are included for reproducibility. 
	\end{abstract}
	
	\section{Introduction and literature review}~\label{sec:literature}
	
	In dynamic infrastructure sciences, the sensor placement (SP) problem is concerned with the time-varying selection or one-time placement of sensors, while optimizing desired objective functions. This problem exists widely in dynamic networks such as transportation systems, electric power systems, and water distribution networks (WDN). The optimal placement of water quality (WQ) sensors is a crucial issue in WDN due to the dangers brought by accidental or intentional contamination, the expensiveness of sensors and their installation cost, and their potential in performing real-time feedback control of water quality---control that requires high-frequency WQ sensor data.
	
	WQ sensor placement in WDN serves different purposes. The high-level one is minimizing the potential public health impacts of a contamination incident given a limited number of sensors.
	To quantify this high-level objective, the WQ literature considers various mathematical objectives and metrics. Specifically, the SP problem has been studied in~\cite{Krause2008,ostfeld2008battle,preis2008multiobjective,schal2013water,Eliades2014,Shastri2006,Aral2010} considering different contamination risks, optimization objectives, optimization formulations, uncertainty, the solution methodology and its computational feasibility, and the use of mobile sensors. Rathi and Gupta~\cite{Rathi2014}  classify methodologies from over forty studies into two categories as single- and multi-objective SP problem. Two other comprehensive surveys  focusing on optimization strategies are also conducted in~\cite{Hart2010,Hu2018}. 
	
	As we mentioned, the most common objective of sensor placement in WDN is to minimize the potential public health caused by contamination incident, and it can be formulated as maximizing the coverage of water with a minimum number of sensors. Lee and Deininger introduce the concept of “Demand Coverage`` and solve the problem using a mixed integer programming (MIP) method~\cite{lee1992optimal}. Kumar et al.~\cite{kumar1997identification} and Kansal et al. \cite{kansal2012identification} propose heuristic methods to find optimal sensor location one by one by selecting one optimal location first and then selecting the next location by modifying the coverage matrix. To consider nodes with lower water quality, Woo et al. modify the objective by placing weights for each term and normalizing the concentrations~\cite{woo2001optimal}. Alzahrani et al. \cite{al2003optimizing} and Afshar and Marino~\cite{afshar2012multi} use genetic algorithm (GA) and ant colony optimization (ACO) respectively to find optimal placement strategy to maximize the demand coverage.  Ghimire et al.~\cite{Ghimire2006} and Rathi and Gupta~\cite{rathi2014locations} also suggested heuristic methods to solve the problem.
	
	We briefly summarize the more recent literature on this problem followed by identifying the key research gap. Recently, He \textit{et al.}~\cite{He2018} propose  a multi-objective SP method to explicitly account for contamination probability variations. Hooshmand \textit{et al.}~\cite{Hooshmand2020} address SP problem with the identification criterion assuming that a limited sensor budget is available, followed by minimizing the number of vulnerable nodes using mixed integer programming (MIP). A combined management strategy for monitoring WDN is proposed in~\cite{Ciaponi2019} based on the application of water network partitioning and the installation of WQ sensors. Winter \textit{et al.}~\cite{Winter2019} investigate optimal sensor placements by introducing two greedy algorithms in which the imperfection of sensors and multiple objectives are taken into account. Giudicianni \textit{et al.}~\cite{Giudicianni2020} present a method that relies on a priori clustering of the WDN and on the installation of WQ sensors at the most central nodes of each cluster---selected according to different topological centrality metrics. Hu \textit{et al.}~\cite{Hu2020}  propose a customized genetic algorithm to solve multi-objective SP in WDN. Based on graph spectral techniques that take advantage on spectrum properties of the adjacency matrix of  WDN graph, a sensor placement strategy is discussed in Di Nardo \textit{et al.}~\cite{DiNardo2018}. Different objective functions leads to different placement strategies, and Tinelli \textit{et al.}~\cite{tinelli2018impact} discuss the impact of objective function selection on optimal placement of sensors. Zhang \textit{et al.}~\cite{zhang2020assessing} investigate the global resilience considering all likely sensor failures that have been rarely explored.
	
	The research community thoroughly investigated water quality sensor placement strategies considering various socio-technical objectives (as briefly discussed above). The objective of this paper is \textit{not} to develop a computational method to solve such SP problems with the aforementioned metrics/objectives. The objective herein is to find optimal SP of water quality sensors considering an overlooked, yet significant metric: the state observability and estimation metric jointly with Kalman filtering.\footnote{In dynamic systems, the Kalman filter is a widely used algorithm that computes unmeasured  state estimates of a system given a dynamic model and data from sensor measurements subject to noise.}   In short, this metric maps sensor placements given a fixed hydraulic profile to a scalar value to be minimized. This value quantifies the observability of unmeasured WQ states (i.e., concentrations of chlorine) in the entire water network. The observability quantification metric is depicted as a state estimation error measuring the difference between the actual WQ states and their estimates. Accordingly, this proposed metric finds the optimal WQ sensor placement that minimizes the state estimation error via the vintage Kalman filter for noisy WQ dynamics and measurement models.

	To the best of our knowledge, this is the first attempt to find the optimal sensor placement jointly with optimizing Kalman filter performance for WQ dynamics. The most related research is the \textit{ensemble Kalman filter}-based techniques by Rajakumar \textit{et al.}~\cite{Rajakumar2019}, where the authors explore the impact of sensor placement on the final state estimation performance. However, the study \textit{(i)}  does not provide sensor placement strategy, \textit{(ii)} mainly focuses on estimating  water quality states and reaction parameters, and \textit{(iii)} a dynamic model for WQ is not present to guide optimal SP. To that end, the objective of this study is to provide a control- and network-theoretic method that determines the optimal geographic placements of water quality sensors while optimizing the Kalman filter performance. The specific paper contributions are:
	\begin{itemize}
		\item The state-space, control-theoretic dynamics depicting the evolution of WQ states, i.e., concentrations of chlorine,  are shown. Specifically, we are modeling and tracking chlorine concentrations as a surrogate for contamination---this has been showcased in various studies depicting rapid depletion of chlorine upon the introduction of contaminants~\cite{yang2008modeling}.
		The dynamics of chlorine concentrations are represented as a time-varying state-space model. This model is then utilized to formulate the water quality sensor placement (WQSP) problem that optimizes the Kalman filter state estimation performance. This formulation \textit{(i)} takes into account and builds a mapping between a WDN observability metric and the performance of Kalman filter and \textit{(ii)} is a set function optimization (an optimization problem that takes sets as variables) that is difficult to solve for large networks.
		\item To account for the time-varying nature of the dynamic WQ model (due to the changes in the hydraulic profiles that are caused by demand profiles),  an algorithm that computes a sensor placement for the most common hydraulic profiles is presented. Furthermore, scalability of this algorithm is investigated. The algorithm is based on an important theoretical feature for set function optimization called submodularity. This feature has been identified in recent control-theoretic studies for sensor placement strategies~\cite{Tzoumas2016,Zhang2017}. In particular, the developed approach is based on a greedy algorithm which returns a suboptimal placement strategy with guarantees on the distance to optimality. An efficient implementation of the algorithm is also presented. Compared to~\cite{Tzoumas2016,Zhang2017}, the proposed algorithm takes into account the time-varying nature of WQ dynamics.
		\item Thorough case studies on three water distribution networks under different conditions are presented. The case studies consider varying scales of water networks, significant demand variability, different number of allocated sensors, and their impact on the state estimation performance and WQSP solution.    Important observations and recommendations for water system operators are given. Github codes are included for reproducibility. 
	\end{itemize}
	The rest of the paper is organized as follows. Section~\ref{sec:Ctrl-WQM} introduces network-oriented water quality dynamic model by presenting the models of each component in detail.  An abstract, linear, state-space format for the water quality model is given first considering the first-order reaction model with known reaction rate coefficients. 
	WQ  observability and its metric (observability Gramian) are introduced in Section~\ref{sec:WQSP}, and then WQSP problem is  formulated and solved by taking advantage of  submodularity property of set function optimization in Section~\ref{sec:WQSPformuation}. A scalable implementation of the problem is showcased.  Section~\ref{sec:test} presents case studies to support the computational algorithms. Appendix~\ref{sec:appa} outlines components of the scalable implementation of the presented computational methods. The notation for this paper is introduced next.
	
	\vspace{1em}
	\noindent \textit{\textbf{Paper's Notation}} $\;$ Italicized, boldface upper and lower case characters represent matrices and column vectors: $a$ is a scalar, $\m a$ is a vector, and $\m A$ is a matrix. Matrix $\m I_n$ denotes a identity square matrix of dimension $n$-by-$n$, whereas $\m 0_{m \times n}$ denotes a zero matrix  with size $m$-by-$n$.
	The notations $\mathbb{R}$ and $\mathbb{R}_{++}$ denote the set of  real and positive real numbers.  The notations $\mathbb{R}^n$ and $\mathbb{R}^{m\times n}$ denote a column vector with $n$ elements and an $m$-by-$n$ matrix in $\mathbb{R}$. 
	For any two matrices $\m A$ and $\m B$ with same number of columns, the notation $\{\m A, \m B\}$ denotes $[\m A^\top \  \m B^\top]^\top$. For a random variable $\m x \in \mathbb{R}^n$, $\mathbb{E}(\m x)$ is its expected value, and its covariance is denoted by $\mathbb{C}(\m x) = \mathbb{E}\left( (\m x - \mathbb{E}(\m x))(\m x - \mathbb{E}(\m x))^\top \right)$.
	
	\section{State-Space Water Quality Dynamic Model}~\label{sec:Ctrl-WQM}
	We model WDN by a directed graph $\mathcal{G} = (\mathcal{W},\mathcal{L})$.  Set $\mathcal{W}$ defines the nodes and is partitioned as $\mathcal{W} = \mathcal{J} \bigcup \mathcal{T} \bigcup \mathcal{R}$ where $\mathcal{J}$, $\mathcal{T}$, and $\mathcal{R}$ are collection of junctions, tanks, and reservoirs. For the $i$-th node, set $\mathcal{N}_i$ collects its neighboring nodes (any two nodes connected by a link) and is partitioned as $\mathcal{N}_i = \mathcal{N}_i^\mathrm{in} \bigcup \mathcal{N}_i^\mathrm{out}$, where $\mathcal{N}_i^\mathrm{in}$ and $\mathcal{N}_i^\mathrm{out}$ are collection of inflow and outflow nodes.  Let $\mathcal{L} \subseteq \mathcal{W} \times \mathcal{W}$ be the set of links, and define the partition $\mathcal{L} = \mathcal{P} \bigcup \mathcal{M} \bigcup \mathcal{V}$, where $\mathcal{P}$, $\mathcal{M}$, and $\mathcal{V}$ represent the collection of pipes, pumps, and valves. In this paper, we the use Lax-Wendroff scheme~\cite{lax1964difference} to space-discretize pipes and each pipe with length $L$ is split into $s_{L}$ segments. 
	%Note that a valve is viewed as a single segment of a pipe, and not considered as a component.  
	The number of junctions, reservoirs, tanks, pipes, pumps and valves is denoted as $n_{\mathrm{J}}$, $n_{\mathrm{R}}$, $n_{\mathrm{TK}}$, $n_{\mathrm{P}}$, $n_{\mathrm{M}}$, and $n_{\mathrm{V}}$. Hence, the number of nodes and links are $n_\mathrm{N} = n_{\mathrm{J}}+n_{\mathrm{R}}+n_{\mathrm{TK}}$ and $n_\mathrm{L} = n_{\mathrm{P}} \cdot s_{L} +n_{\mathrm{M}}+n_{\mathrm{V}}$.
	
	The principal component of the presented state-space, control-theoretic water quality modeling is a state-vector defining the concentrations of the disinfectant (chlorine) in the network. Concentrations at nodes such as junctions, reservoirs, and tanks are collected in vector $\m c_\mathrm{N} \triangleq \{\m c_\mathrm{J}, \m c_\mathrm{R}, \m c_\mathrm{T} \} $; concentrations at links such as pipes and pumps are collected in $\m c_\mathrm{L} \triangleq \{\m c_\mathrm{P}, \m c_\mathrm{M}, \m c_\mathrm{V} \} $. We define WQ state $\m x(t) \triangleq \m x$ at time $t$ as:
	%\begin{equation*}
	$\m x(t) =  \{\m c_\mathrm{N},\m c_\mathrm{L} \}  =  \{\m c_\mathrm{J}, \m c_\mathrm{R}, \m c_\mathrm{T},  \m c_\mathrm{P}, \m c_\mathrm{M}, \m c_\mathrm{V}\} \in \mbb{R}^{n_x}, n_x = n_\mathrm{N}  + n_\mathrm{L}. $
	
	%\end{equation*}
	We also make two assumptions: \textit{(i)}  the mixing of the solute is complete and instantaneous  at junctions and  in tanks with a continuously stirred tank reactors (CSTR) model~\cite{rossman2000epanet}, and \textit{(ii)} the first-order reaction for single-species that describes disinfectant decay both in the bulk flow and at the pipe wall are assumed herein.
	% , and \textit{(iii)} the upstream nodes of pumps are only reservoirs which are the source of solute and their concentration remains constant. 
	The assumptions are widely used in the literature~\cite{rossman1996numerical,basha2007eulerian,shang2008epanet}.
	%  and the third is made for brevity of derivations and has no implication on the correctness of the paper's presented developments. 

	\subsection{Conservation of mass}\label{sec:conservation}
	
	The water quality model represents the movement of all chemical and/or microbial species (contaminant, disinfectants, DBPs, metals, etc.) within a WDN as they traverse various components of the network. Specifically, we are considering the single-species interaction and dynamics of  chlorine. This movement or time-evolution is based on three principles: \textit{(i)} \textit{mass balance in pipes}, which is represented by chlorine transport in differential pipe lengths by advection in addition to its decay/growth due to reactions; \textit{(ii)} \textit{mass balance at junctions}, which is represented by complete and instantaneous mixing of all inflows, that is the concentration of chlorine in links flowing into this junction; and \textit{(iii)} \textit{mass balance in tanks}, which is represented by a continuously stirred tank reactors (CSTRs)~\cite{rossman2000epanet} model with complete and instantaneous mixing and  growth/decay reactions. The modeling of each component is introduced next.   
	
	\subsubsection{Chlorine transport and reaction in pipes}  The water quality modeling for pipes involves modeling the chlorine transport and reaction by  1-D advection-reaction (A-R) equation. For any Pipe $i$, the 1-D A-R model is given by a PDE:
	\begin{equation} ~\label{equ:adv-reac}
		{\partial_t c_\mathrm{P}} =  -v_{i}(t) {\partial_x c_\mathrm{P}} + r_{i} c_\mathrm{P} ,
	\end{equation}
	\noindent where $v_{i}(t)$ is  flow velocity, $r_{i}$ is the first-order reaction rate and remains constant, which is related with the bulk and wall reaction rate and mass transfer coefficient between the bulk flow and the pipe wall~\cite{basha2007eulerian,rossman1996numerical}.  Here, the Lax-Wendroff (L-W) scheme~\cite{lax1964difference} shown in Fig.~\ref{fig:lax} is used to approximate the solution of the PDE~\eqref{equ:adv-reac} in space and time; this
	model has been used and accepted in the literature~\cite{rossman1996numerical,morais2012fast,fabrie2010quality}. Pipe $i$ with length $L_{i}$ is split into $s_{L_{i}}$ segments, and  the discretized form  for segment $s$ is given by
	\begin{equation}~\label{equ:adv-reac-lax}
		\hspace{-1em}    c_{i,s}(t\hspace{-1pt}+\hspace{-1pt}\Delta t) = \underline{\alpha}    c_{i,s-1}(t) 
		+ (\alpha + r_i)c_{i,s}(t) +\bar{\alpha} c_{i,s+1}(t),
	\end{equation}
	where L-W coefficients for previous, current, and next segment  are $\underline{\alpha} = 0.5 \beta (1+\beta)$, ${\alpha} =  1- \beta^2 $, and $\bar{\alpha} = -0.5 \beta (1-\beta)$. Note that $\beta \in \left(0,1\right]$ for Pipe $i$ at time $t$ is a constant related with stability condition of  L-W scheme, and can be decided by ${v_{i}(t) \Delta t}(\Delta x_{i})^{-1}$, where $\Delta t$ and $\Delta x_{i}$ are the time step and the space-discretization step in Fig.~\ref{fig:lax}. Hence, to stabilize L-W scheme, the water quality time step  $\Delta t \leq \min({\Delta x_{i}}/v_{i}(t))$, for all $i \in \mathcal{P}$. The L-W scheme coefficients $\underline{\alpha}$, $\alpha$, and $\bar{\alpha}$ are a function of time but vary much slower than $\m x(t)$, and they only change when $v_i(t)$ changes after the $\Delta t$ and $\Delta x_i$ are fixed. That is, they only update each hydraulic time step. Equation~\eqref{equ:adv-reac-lax} can be lumped in a matrix-vector form for all segments $s$ for all Pipes $i \in \mc{P}$ as:
	\begin{figure}[t]
		\centering
		\includegraphics[width=1\linewidth]{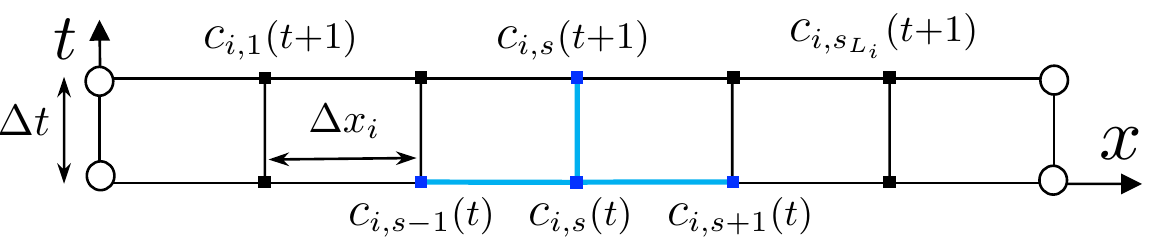}
		\caption{Time-space discretization of Pipe $i$ based on the L-W scheme. }
		\label{fig:lax}
		\vspace{-1.em}
	\end{figure}
	\begin{equation}~\label{equ:abs-pipe-mass}
		\m c_\mathrm{P}(t+ \Delta t) =  \m A_\mathrm{P}(t) \m c_\mathrm{P}(t)  + \m A_\mathrm{N}(t) \m c_\mathrm{N}(t),
	\end{equation}
	where matrices $\m A_\mathrm{P}$ and  $\m A_\mathrm{N}$ map the scalar equation~\eqref{equ:adv-reac-lax} into the vector form~\eqref{equ:abs-pipe-mass}. The Github codes of this paper~\cite{wang_2020} shows how these matrices are computed.
	
	\subsubsection{Chlorine mass balance at junctions}~\label{sec:mbjunction} Mass conservation of the disinfectant (i.e., chlorine) for Junction $i$ at time $t$ can be described by
	\begin{equation} 
		\hspace{-1.5em}
		\textstyle \sum_{k = 1}^{|\mathcal{N}_i^\mathrm{in}|} q_{ki}(t)c_{ki}(t)  
		=
		d_i(t) c_{i}(t) + \textstyle \sum_{j = 1}^{|\mathcal{N}_i^\mathrm{out}|} q_{ij}(t)  c_{ij}(t),~\notag
	\end{equation}
	where $ \{ ki : k \in \mathcal{N}_i^\mathrm{in} \} $ and  $ \{ ij : j \in \mathcal{N}_i^\mathrm{out} \} $ represent the sets of links with inflows and outflows of Junction $i$; $d_i$ is its demand; $q_{ki}(t)$ and $q_{ij}(t)$ are the flow rate in Links $ki$ and $ij$; $c_{ki}(t)$ and $c_{ij}(t)$ are the corresponding concentrations. Specifically, when links are pipes, $c_{ki}(t)$ and $c_{ij}(t)$ should be the last and first segment of Pipes $ki$ and $ij$. The matrix form when considering all junctions is given as
	\begin{equation} ~\label{equ:abs-junction-mass}
		\m c_\mathrm{J}(t+ \Delta t) =  \m A_\mathrm{J}(t) \m c_\mathrm{J}(t)  + \m A_\mathrm{L}(t) \m c_\mathrm{L}(t).
	\end{equation}
	% where $\m A_\mathrm{J}$ is self-contribution matrix for junctions, and $\m A_\mathrm{L}$ is the contribution matrix from links.
	\subsubsection{Chlorine mass balance at tanks}  
	Akin to dealing with junctions, we can express the mass balance equations for each tank, the details are similar and omitted for brevity and ease of exposition. With that in mind, the provided Github codes present all of the necessary details that are required to arrive at the high-fidelity state-space description. We directly give the matrix form of all tanks as
	\begin{equation} ~\label{equ:abs-tank-mass}
		\m c_\mathrm{T}(t+ \Delta t) =  \m A_\mathrm{T}(t) \m c_\mathrm{T}(t)  + \m A'_\mathrm{P}(t) \m c_\mathrm{P}(t),
	\end{equation}
	where $\m A_\mathrm{T}$ is in terms of tank volumes $\m V_\mathrm{T}$, time step $\Delta t$, and flow rates flowing in or out of tanks. 
	%It is the self-contribution matrix for tanks, and $\m A'_\mathrm{P}$ is the contribution matrix from pipes (since tanks always connecting to pipes).

	\subsubsection{Chlorine mass balance at reservoirs} 
	Without loss of generality, it is assumed that the chlorine sources are only located at reservoirs, and the concentration at a reservoir is constant. That is, 
	\begin{equation}\label{equ:abs-reservoir}
		\m c_\mathrm{R}(t + \Delta t)  = \m c_\mathrm{R}(t) .
	\end{equation}
	
	\subsubsection{Chlorine transport in pumps and valves}~\label{sec:PumpandValve} 
	We consider that the \textit{lengths} of pumps to be null, i.e., the distance between its upstream node and downstream node is zero, and hence they neither store any water nor are discretized into different segments. Therefore, the concentrations at pumps or valves equal the concentration of upstream nodes (a reservoir) they are connecting. That is
	\begin{equation}
		\hspace{-1em} c_{j}(t+\Delta t) =  c_i(t + \Delta t) =  c_i(t) =  c_j(t), i \in \mathcal{R}, j \in \mathcal{M},~\notag
	\end{equation}
	and the corresponding matrix form for pumps  is 
	\begin{equation}~\label{equ:abs-pump}
		\m c_\mathrm{M}(t+\Delta t) = \m c_\mathrm{M}(t). 
	\end{equation}
	As for valves installed on pipes, it is simply treated as a segment of that pipe. In this case, the concentration in valves equals the segment concentrations in pipes.  
	We next show how these matrix forms can yield state-space formulation of water quality modeling.
	\subsection{Water quality modeling in state-space form}~\label{sec:space-state}
\begin{figure}[h]
		\begin{equation}~\label{equ:de-abstract1}
			\hspace{-2em}\includegraphics[width=0.92\linewidth,valign=c]{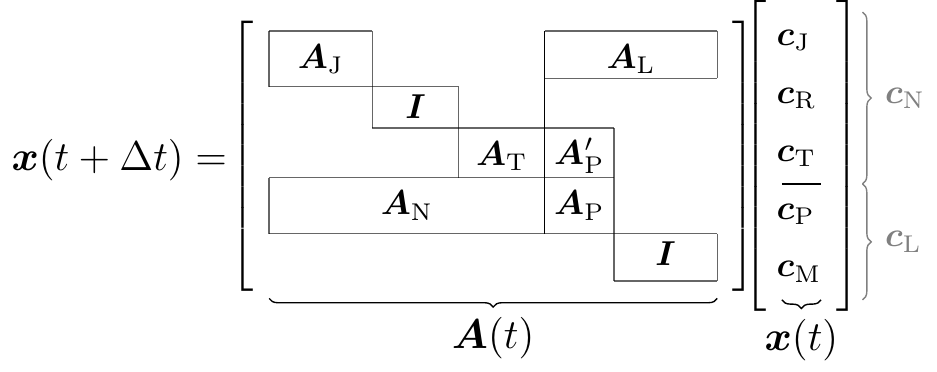}.
		\end{equation}
	\end{figure}

	The briefly summarized water quality model of each component from the previous section
	% Section~\ref{sec:conservation} 
	can be written as a state-space Linear Difference Equation (LDE) as in~\eqref{equ:de-abstract1} where $\m I$ is an identity matrix of appropriate dimension.
	For the ease of exposition, we consider that $\Delta t = 1 \sec$ and the time-index $t$ is replaced with another time-index $k$. The state-space form of the water quality model is presented as a linear time-variant (LTV) system:
	\begin{equation}~\label{equ:ltv}
		{\m{x}}(k+1) =\m A(k) \m{x}(k)+\m  {w}(k), \;\; \m y (k)=\m C \m{x}(k)+ \m{v}(k),
	\end{equation}
	where $\m x(k) \in \mbb{R}^{n_x}$ is the state vector defined above; $\m y(k) \in \mbb{R}^{n_y}$ represents a vector of data from WQ sensors; $\m {w}(k) \in \mbb{R}^{n_x}$ and $\m{v}(k) \in \mbb{R}^{n_y}$ are the process and measurement noise; $\m C \in \mbb{R}^{n_y \times n_x}$ is a matrix depicting the location of the placed sensors where $n_y << n_x$.
	% and each row has at most one element which is 1, since sensors is only installed at nodes. s
	We note the following. First, although the argument of the state-space matrices  $\m A(k)$  is in terms of $k$, this is somewhat of an abuse for the notation seeing that $\m A(k)$ encodes the hydraulic profile (heads and flow rates) that does not change with the same frequency as the water quality states $\m x(k)$. Hence, the state-space model~\eqref{equ:ltv} is time varying as system matrix $\m A(k)$ changes for different hydraulic simulation, but remains the same $\m A$ in a single simulation. Second, and without loss of generality, the input vector from booster stations is implicitly embedded within the state-space matrix $\m A$. Third, for all $k \geq 0$, it is assumed that initial condition, process noise $\m  {w}(k)$ and the measurement noise $\m v(k)$ are uncorrelated and the noise variance from each sensor is $\sigma^2$. Finally, we like to point out that extensive details for the above state-space model can be studied from our recent work on model predictive control of water quality dynamics~\cite{wang2020effective}.
	\section{Observability Metrics for WQ Dynamics}~\label{sec:WQSP}
	The objective of this section is two-fold. First, to introduce water system engineers and researchers to control-theoretic approaches for ensuring or optimizing the observability of the water quality dynamics. Herein, observability is defined as the ability to estimate water quality model states $\m x(k)$ from available measurements $\m y(k)$ via a state estimation routine. This provides situational awareness for the operator given data from few water quality sensors. Second, to define a simple observability metric that maps the number and location of fixed sensors to a scalar metric acting as a proxy for the state estimation. 
	%The majority of the ensuing discussions in this section are guided by the recent theoretical developments in the control theoretic literature. An objective of the paper is hence to apply and adapt some of these developments for the water quality sensor placement problem.
	\subsection{Metrics for observability and its interpretations}~\label{sec:metric}
	In dynamic systems theory, observability is a measure of how the system state vector $\m{x}(k) \in \mbb{R}^{n_x}$ can be inferred from knowledge of its output $\m{y}(k) \in \mbb{R}^{n_y}$ over either finite- or infinite-time horizons. In particular, given sensor data $\m y(0), \m y(1), \ldots, \m y(k_f-1)$ for finite $k_{f} = k_{final}$ time-steps, observability is concerned with reconstructing or estimating the initial unknown state vector $\m x(0)$ from the $k_f$ measurements, and subsequently computing $\m x(1), \ldots, \m x(k_f)$ assuming noiseless system.  Accordingly, a linear dynamic system (such as the water quality model~\eqref{equ:ltv}) is observable if and only if the observability matrix for $k_f$ time-steps 
	\begin{equation}
		\mathcal{O}(k_f) = \{\m C, \m C \m A, \hdots, \m C \m A^{k_f-1}\} \in \mbb{R}^{k_f n_y \times  n_x }~\notag
	\end{equation}
	is full column rank~\cite{hespanha2018linear}, i.e., $\rank(\mc{O}(k_f))=n_x$ assuming that $k_fn_y > n_x$. In this section, and for brevity, we assume that the hydraulic variables are not changing during each hydraulic simulation period and hence $\m A(k)= \m A$. With that in mind, the proposed sensor placement formulations considers changing hydraulic simulations. 
	
	For the infinite-time horizon case with $k_f = \infty$ (that is, data has been collected over a long period of time), a system is observable if and only if the observability matrix $\mathcal{O}(k_f=n_x) \in \mbb{R}^{n_x n_y \times  n_x }$ is full column rank~\cite{hespanha2018linear}.  However, observability is a binary metric---it cannot indicate \textit{how observable} a dynamic system is. Due to the complexity and dimension of the water quality model~\eqref{equ:ltv}, this dynamic model is \textit{not} observable, i.e., it fails the aforementioned rank condition for various water networks and hydraulic simulation profiles. Specifically, it is virtually impossible to accurately reconstruct all chlorine concentrations (states $\m x(k)$) unless water quality sensors are ubiquitously available and widespread in the network, i.e., installed at each junction. 
	
	To that end, a more elaborate, non-binary quantitative metric for observability  is needed for the water quality model and the sensor placement problem. One metric is based on the \textit{observability Gramian}~\cite{hespanha2018linear}
	defined as the $k_f$ sum of matrices
	\begin{equation}
		\m W(k_f)=\sum_{\tau=0}^{k_f}\left(\m A^{\top}\right)^{\tau} \m C^{\top} \m C \m A^{\tau}.~\notag
	\end{equation}
	The system is observable at time-step $k_f$ if matrix $\m W(k_f)$  is nonsingular and is unobservable if $\m W(k_f)$ is singular. Similarly, this definition extends for the infinite-horizon case with $k_f = \infty$. However, $\m W$ is still a matrix and the aforementioned observability-singularity discussion is still binary. As a result, various non-binary metrics have been explored in the literature~\cite{Summers2013,Summers2016}. This includes: the minimum eigenvalue $\lambda_{\mathrm{min}}(\m W)$, the log determinant $\log \operatorname{det} (\m W)$, the $\operatorname{trace} (\m W)$, and the sums or products of the first $m$ eigenvalues $\lambda_1,\hdots,\lambda_m$ of $\m W$.  These metrics differ in their practical application, interpretation, and theoretical properties; the reader is referred to~\cite{Summers2016} for a thorough discussion. In this paper, we utilize the $\log \operatorname{det} (\m W)$ metric due to various reasons outlined in the ensuing sections, but the formulations presented in the paper can be extend to other metrics. 
	
	\subsection{Metrics for water quality observability matrix}~\label{sec:aug}
	In this section, we provide a discussion on the utilized metric for observability for the sensor placement problem. To do so, we consider the time-invariant state-space matrices for a single hydraulic simulation $k \in [0,k_f]$ which is also a single instant of hydraulic simulation and demand profile. That is, to ease the ensuing exposition we assume that the state-space matrix $\m A(k) = \m A$ is fixed rather than being time-varying (the actual methods consider time-varying demand pattern). The objective of this section is to formulate a water quality observability metric that maps collection of water quality data $\m y(k)$ from a specific number of sensors $n_y$ to a scalar observability measure under the noise from water quality dynamics and measurement models.

	First, consider the augmented measurement  vector $\bar{\m y}(k_f) \triangleq  \{\m y(0), \ldots,\m y(k_f) \}$ for $k_f+1$ time-steps. Given~\eqref{equ:ltv}, this yields:
	\begin{equation}~\label{equ:ymeasurement}
		\hspace{-1.47em}\includegraphics[width=0.92\linewidth,valign=c]{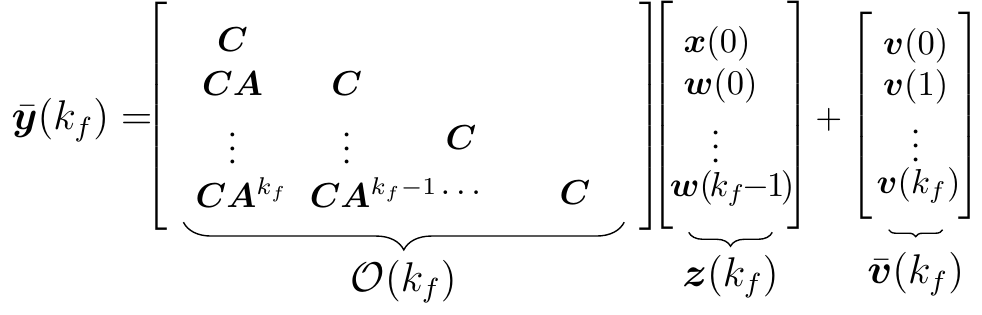},
	\end{equation}
	where $\m z(k_f)$ lumps initial unknown state $\m x_0 = \m x(0)$ and process noise $\m w(k_f)$, and $\bar{\m v}(k_f)$ collects all measurement noise. Note that the left-hand side of~\eqref{equ:ymeasurement} is known, whereas vectors $\m z(k_f)$ and $\bar{\m v}(k_f)$ are unknown vectors. To that end, the problem of estimating $\m z(k_f) \triangleq \m z \in \mbb{R}^{n_z}$, where $n_z= (k_f+1)n_x$, is important to gain network-wide observability of water quality state $\m x$ which will guide the real-time estimation. As a probabilistic surrogate to estimating $\m z$,  we utilize the minimum mean square estimate (MMSE) defined as
	$\mbb{E}(\m z - \hat{\m z})$, and its corresponding posterior error covariance matrix $\Sigma_{\m z}$. These two quantities provide estimates of means and variances of the unknown variable $\m z(k_f)$. Interestingly, these can be written in terms of the sensor noise variance $\sigma^2$, the collected sensor data $\bar{\m y}(k_f)$, the observability-like matrix  $\mathcal{\m O}(k_f)$ in~\eqref{equ:ymeasurement}, and the expectation and covariance of the unknown variable $\m z(k_f)$ given by
	\begin{equation}
		\mathbb{E}(\m z(k_f)), \; \mathbb{C}(\m z(k_f)) =  \mathbb{E}\left( (\m z - \mathbb{E}(\m z))(\m z - \mathbb{E}(\m z))^\top \right) ~\notag
	\end{equation}
	Given these developments, and to guide the sensor placement problem formulation, a metric is needed to map the covariance matrix $\Sigma_{\m z}$ to a \textit{scalar} value. In particular, the metric $\log \operatorname{det}\left(\Sigma_{\m z}\right)$, which maps an $n_z$-by-$n_z$ matrix $\Sigma_z$ to a scalar value, can be used to achieve that. Fortunately, $\log \det (\Sigma_{\m z})$ has a closed form expression given by:
	\begin{equation}~\label{equ:closedform}
		\hspace{-0.352cm}\log \det (\Sigma_{\m z}) =   2 n_z \log (\sigma)\hspace{-2pt}-\hspace{-2pt}\log \det \left(\sigma^{2} \mathbb{C}^{-1}\left(\m z\right)+\m W_o \right)
	\end{equation} 
	where $\m W_o = \mathcal{\m O}^{\top}(k_f)\mathcal{\m O}(k_f)$. The reader is referred to~\cite{Tzoumas2016}  for the derivation of~\eqref{equ:closedform}. We note the following: \textit{(i)} the closed-form expression of $\log \operatorname{det}\left(\Sigma_{\m z}\right)$ in~\eqref{equ:closedform} assumes a \textit{fixed} sensor placement while associating a scalar measure of water quality observability given a collection of sensor data and the system's parameters. This closed form expression is rather too complex to be incorporated within a sensor placement formulation and does not allow for near real-time state estimation. The next section discusses simple solutions to these issues.  \textit{(ii)} We use the $\log \det (\cdot)$ metric here as it is endowed with desirable theoretical properties (namely super/sub-modularity) that makes it amenable to large-scale networks, it exhibits a closed-form expression as in~\eqref{equ:closedform}, and has been used in various sensor placement studies in the literature. With that in mind, other metrics can be used including the $\mathrm{trace}$ operator. 
	\subsection{Relationship with the Kalman filter}
	The above discussions yield a metric that can be used for quantifying observability of the water quality model~\eqref{equ:ltv}, in addition to probabilistically estimating the unknown, initial state vector $\m x(0)$. A relevant problem is the real-time state estimation via the Kalman filter, which essentially reconstructs or estimates in real-time states $\m x(k)$ from output measurements $\m y(k)$. This is in contrast with the batch state estimation as in~\eqref{equ:ymeasurement}. While the initial state estimation problem discussed in the previous section provides a starting point for reconstructing $\m x$, the Kalman filter presents a more general approach to the estimation problem. In fact, ignoring the process noise $\m w$ and setting variances of sensor data to $\sigma^2 = 1$, the Kalman filter becomes equivalent to a real-time version of the above probabilistic estimator. Most importantly, the metric $\log \operatorname{det} (\cdot)$ degenerates to:
	\begin{align}
		\hspace{-0.4cm} \log \operatorname{det}\left(\Sigma_{\m z}\right) &=   - \log \operatorname{det} ( \m I_{n_{x}} + \m W(k_f) )  ~\label{equ:obsmetric} \\
		&=  - \log \operatorname{det} \left(  \m I_{n_{x}} + \sum_{\tau=0}^{k_f}\left(\m A^{\top}\right)^{\tau} \m C^{\top} \m C \m A^{\tau}  \right) \notag
	\end{align} 
	where $\m I_{n_{x}}$ is an identity matrix of size $n_x$. This is shown in the recent control theoretic literature~\cite{Jawaid2015,Zhang2017}. In short, this is a simple metric that maps the number of installed or placed sensors (i.e., number of rows of matrix $\m C$) to a metric that defines the quality of the state estimates. When no sensor is installed or $\m C$ is a zero matrix, the observability Gramian $\m W(k_f)$ is also a zero matrix, and intuitively the $\log \det (\cdot)$ metric defined above has the maximum error of $0$. When the network is fully sensed---that is $n_y = n_x$, $\m C = \m I_{n_{x}}$, and all states are measured---then $\m W(k_f) = \m I_{n_x} + \m A + \hdots + \m A^{k_f}$ and the smallest error is achieved. 
	
	Building on that, the control theoretic literature thoroughly investigated bounds for the estimation error and the corresponding metric with respect to the number of sensors; see~\cite[Theorem 1]{Tzoumas2016}. The objective of this paper is to build on these developments and investigate how such metric relates with the performance of the Kalman filter. The next section formulates the water quality sensor problem using the introduced metric. 
	
	\section{Water Quality Sensor Placement Formulation}~\label{sec:WQSPformuation}
	%\subsection{WQSP Formulation}
	The objective of the presented water quality sensor placement (WQSP) formulation is to minimize the error covariance of the Kalman filter while using at most $r$ water quality sensors. In WDN, water quality sensors are installed at nodes, that is,  at most $r$ sensors are selected from the set $\mathcal{W} = \mathcal{J} \bigcup \mathcal{T} \bigcup \mathcal{R}$ where the cardinality of set $|\mathcal{W}| = n_N$, i.e., the set $\mc{W}$ contains $n_N$ possible locations at various junctions, tanks, and reservoirs. This forms a sensor set $\mathcal{S} \subset \mathcal{W}$ where $|\mathcal{S}| = n_{\mathcal{S}} \leq r$. The specific geographic placement and locations of these $n_{\mathcal{S}}$ sensors are encoded in matrix $\m C$ of~\eqref{equ:ltv} through binary indicators.  In short, the presented WQSP seeks to find the optimal set $\mc{S}^*_{r}$ that optimizes the state estimation performance with at most $r$ WQ sensors. 
	
	The metric discussed in the previous section assumes that the state-space matrix $\m A$ (encoding network and hydraulic simulation parameters) is time-varying due to varying demand and flow/head profiles. In short, the metric~\eqref{equ:obsmetric} yields a time-varying value and hence different state estimation performance for each hydraulic simulation reflected with a different $\m A(k)$ matrix.  As a result, considering a varying hydraulic simulation profile within the sensor placement problem is important, i.e., the sensor placement solution needs to be {aware} of the most probable demand and hydraulic scenarios. 
	
	Consequently, we define $\m D_i \in\mathbb{R}^{n_{\mathrm{J}} \times T_h k_f}, \forall i \in \{1,\ldots,n_d\}$ for all $n_{\mathrm{J}}$ junctions during $T_h$ distinct hydraulic simulations, each lasting $k_f \, \sec$. The notation $\m D_{i,k}$ defines the $k$th column vector of matrix $\m D_i$. Parameter $n_d$ reflects the number of potential demand patterns; concrete examples are given in case study section. Demand profiles  $\m D_i \in \mathcal{D}$ essentially define the most common varying demand profiles experienced by the system operator from historical data. Each demand profile results in a different hydraulic profile and hence a different state-space matrix\footnote{We defined $\m A(k)$ earlier due to the change in the hydraulic and demand profiles. The notation $\m A(\m D_{i,k})$ is equivalent to $\m A(k)$ but offers more clarity.} $\m A(\m D_{i,k})\triangleq \m A(k)$. Given these definitions and for an a priori defined $\m D_{i,k} \in \mc{D}$, one useful instant of the WQSP problem can be abstractly formulated as:
	\begin{equation}\label{equ:WSQP}
		\begin{split}
			\mathrm{{WQSP:}} \;\;\;\;   \minimize \;\; \; &  f(\mathcal{S}; \m A(\m D_{i,k})) \\
			\subjectto \;\;\;& 
			%\m h = g(\m d_{P_1},\ldots, \m d_{P_N}),  \notag \\
			%& 
			{\mathcal{S} \subset \mathcal{W}, \;\; |\mathcal{S}| = n_\mathcal{S}} \leq r. 
		\end{split}
	\end{equation}
	The design variable in the optimization problem $\mathrm{WQSP}$ is the location of the installed sensors reflected via set $\mc{S}$ defined earlier. The objective function $f(\cdot;\cdot): \mathbb{R}^{n_{\mc{S}}} \times \mathbb{R}^{n_x \times n_x} \to \mathbb{R}$ maps the optimal sensor placement candidate $\mc{S}$ and given hydraulic demand profile $\m D_{i,k}$ and its corresponding matrix $\m A(\m D_{i,k}) $ to the state estimation, Kalman filter performance. We note that when the objective function has a set as the variable (i.e., $\mc{S}$ in $f(\cdot;\cdot)$), the objective function is often referred to as a \textit{set function}. We use these terms interchangeably. 
	In this paper, the set (objective) function takes the form of~\eqref{equ:obsmetric} which indeed takes explicitly the sensor placement set $\mc{S}$ through matrix $\m C$  as well as the a priori known hydraulic profiles and the corresponding state-space matrices  $\m A(\m D_{i,k})$. The constraint set of $\mathrm{WQSP}$ represents the number of utilized sensors and their location in the network.

	For small-scale water networks, one may solve the set function optimization~\eqref{equ:WSQP} via brute force, but this is impossible for large-scale networks---such problems are known to be an NP-hard one, i.e., a computational problem that is suspected to have no polynomial-time algorithm to optimally solve. To address this computational challenge, we resort to a widely-used approach in combinatorial optimization: exploiting special property of the set function $f(\mathcal{S};\m A(\m D_{i,k}))$ via sub/super-modularity defined as follows. 
	
	A set function $f(\cdot)$ is submodular if and only if $ f(\mc{A} \cup\{a\})-f(\mc{A}) \geq f(\mc{B} \cup\{a\})-f(\mc{B})$ for any subsets $\mc{A} \subseteq \mc{B} \subseteq \mc{V}$ and $\{a\} \in \mc{V} \backslash \mc{B}$. A set function $f(\cdot)$ is supermodular if $-f(\cdot)$ is submodular. Intuitively, submodularity is a diminishing returns property where adding an element to a smaller set gives a larger gain than adding one to a larger set~\cite{lovasz1983submodular}.
	
	The computational framework of submodularity of set function optimization allows one to use greedy algorithms~\cite{cormen2009introduction} with desirable performance while being computationally tractable. Although greedy algorithms are known to return suboptimal solutions, they are also known to return excellent performance when the set function is especially sub/super-modular. Interestingly, the set function in $\mathrm{WQSP}$ given in~\eqref{equ:obsmetric} is indeed supermodular~\cite[Theorem 2]{Tzoumas2016}. Given this property, a vintage greedy algorithm---applied to solve the NP-hard problem $\mathrm{WSQP}$---can return a solution $\mc{S}$ with objective function value $f(\mc{S})$ \textit{at least} 63\% of the optimal solution $f(\mathcal{S}^{*})$~\cite{Tzoumas2016}. Empirically, a large body of work~\cite{Tzoumas2016,Zhang2017,Cortesi2014} shows that the solution provided by some greedy algorithms can be near-optimal, rather than being 63\% optimal.
	
	\begin{algorithm}[t]
		\small	\DontPrintSemicolon
		\KwIn{Number of sensors $r$, all demand profiles $\mathcal{D}$, water network parameters, $k = i = 1$, $\tilde{\mc{S}} = \emptyset$}
		\KwOut{Optimal sensor set $\mathcal{S^{\star}}$}
		\textbf{Compute:}  $\m A(\m D_{i,k})=\m A, \forall i, k \in \{1,\ldots,n_d\},  \{1,\ldots, T_h k_f\}$\;
		\For{$k\leq T_hk_f$ }{
			\textcolor{blue}{// \textbf{For each single hydraulic simulation interval} $k$}		\;		 
			$i = 1$, $\bar{\mc{S}}=\emptyset $\;
			\For{$i \leq n_d$ }{
				\textcolor{blue}{// \textbf{For each demand profile}} \;
				$j = 1, \mathcal{S}_j =  \emptyset $\;
				\While {  $j \leq r$ }{
					$e_{j} \leftarrow  \mathrm{argmax} _{e  \in \mc{W} \backslash \mathcal{S} }\left[f(\mathcal{S};\m A )-f(\mathcal{S}  \cup\{e\};\m A)\right]$\;
					$\mathcal{S}_j  \leftarrow \mathcal{S}_j  \cup\left\{e_{j}\right\}$\;
					$j \leftarrow j+1$		
				}
				$\bar{\mc{S}} \leftarrow \bar{\mc{S}} \bigcup {\mc{S}}_{j}$,  $i \leftarrow i+1$
			}
			%					Find the minimum $f(\mathcal{S}_{ik};\m D_{ik})$ at $k$ and the corresponding set $\mathcal{S}_{ik}$ is denoted as $\mathcal{S}^\star_{k}$\;
			$ {\mc{S}}^{(k)} \leftarrow \arg \max_{\mc{S} \in \bar{\mc{S}}} f(\mathcal{S}; \m A)$\;
			$\tilde{\mc{S}} \leftarrow \tilde{\mc{S}} \bigcup {\mc{S}}^{(k)}$ \;
			$k \leftarrow k+k_f$ 
			%					\textcolor{blue}{\textbf{// \textbf{Next hydraulic simulation}} $k$} \;
		}
		$\mathcal{S}^*  \leftarrow \arg \max_{\mc{S} \in \tilde{\mc{S}}} {T}(\mathcal{S})$ 	\textcolor{blue}{\textbf{// \textbf{Greedy-optimal sensor placement}}}\;
		\caption{Greedy algorithm to solve WQSP problem.}
		\label{alg:greedy}
	\end{algorithm}

	We apply a greedy algorithm to solve the WQSP for various hydraulic profiles. The details of this algorithm are given in Algorithm~\ref{alg:greedy}. The notation $\mc{S}_j$ denotes the sensor set with $j$ placed sensors. The notation $\mc{S}^{(k)}$ denotes the sensor set at iteration $k$. The sets $\tilde{\mc{S}}$ and $\bar{\mc{S}}$ are super-sets that include various sets $\mc{S}$. Variable $e \in \mc{S}$ defines an element (i.e., a junction) in the set $\mc{S}$. The inputs for the algorithm are the number of sensors $r$, all demand profiles $\m D_{i,k} \in \mc{D}$, and WDN parameters. The output of the algorithm is greedy-optimal sensor set $\mc{S}^*$. The first step of the algorithm is to compute all state-space matrices $\m A(\m D_{i,k})$ for various demand profiles $\m D_{i,k}$. Then, given a fixed hydraulic simulation interval $k$, a fixed demand profile $i$,  and fixed number of sensors $j$, Step 9 computes the optimal element in the set $\mc{W} \backslash \mc{S}_j$ that yields the best improvement in the set function optimization reflecting the Kalman filter performance---a core component of the greedy algorithm and supermodular optimization. At each iteration inside the while loop, the algorithm finds the optimal element $e_j$ (i.e., the sensor through a junction ID) that results in the best improvement in the state estimation performance metric. 
	
	Then, the $n_d$ sets $\mc{S}_j$ (that include the optimal sensor sets for all $n_d$ demand profiles) are stored in a master set $\bar{\mc{S}}$. This is then followed by finding the optimal sensor sets from $\bar{\mc{S}}$ for all $T_h$ hydraulic simulations; these are all included in another master set $\tilde{\mc{S}}$. Finally, the algorithm terminates by computing the final {optimal sensor locations  $\mc{S}^*$} via picking the combination that maximizes the occupation time ${T}(\mc{S})$ for all $\mc{S} \in \tilde{\mc{S}}$, i.e.,  a metric that defines the frequency of a specific sensor activation. Finally, we note that this algorithm returns the \textit{greedy-optimal} solution. This solution is not necessarily the optimal solution as discussed above with the 63\% optimality guarantees. Thorough case studies are given in the ensuing section.
	\begin{figure}[t]
		\centering
		\includegraphics[width=1\linewidth]{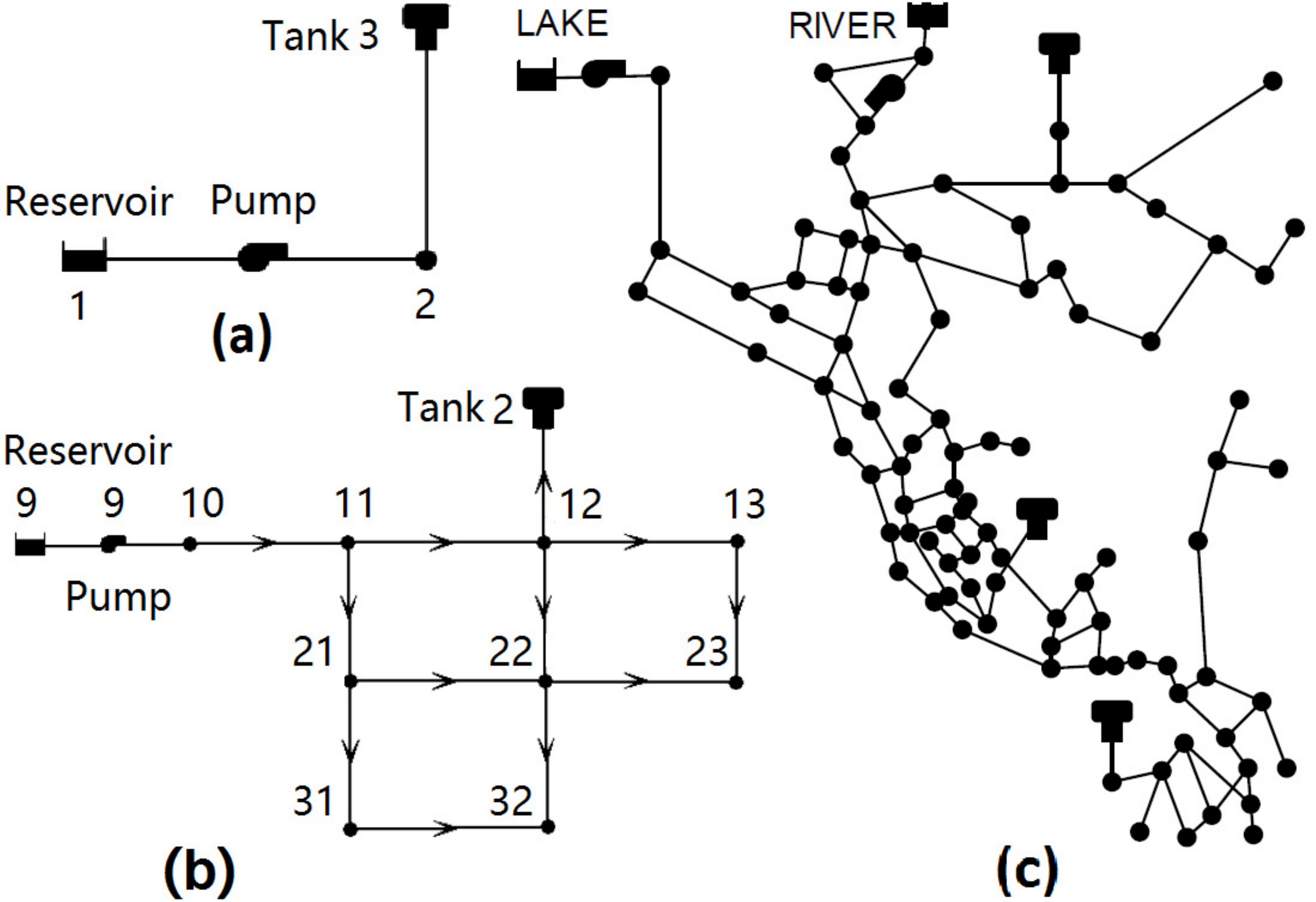}
		\caption{(a) Three-node network, (b) Net1, and (c) Net3.}
		\label{fig:setup}% THIS figure is in the LDE branch on github.
	\end{figure} 
	\section{Case Studies}~\label{sec:test}  
	We present three simulation examples (three-node network, Net1, and Net3 network~\cite{rossman2000epanet,shang2002particle}) to illustrate the applicability of our approach. The three-node network is designed to illustrate the details of proposed method and help readers understand the physical meaning of results intuitively. Then, we test Net1 with looped network topology considering the impacts on final WQSP from choosing \textit{(i)} the length of a single hydraulic simulation $t$, \textit{(ii)} L-W scheme time-step $\Delta t$ (or equally dynamic number of segments), \textit{(iii)} different base demands, and \textit{(iv)} different patterns. Net3 network is used to test scalability of proposed algorithm and verify our findings further. Considering that the LDE model~\eqref{equ:ltv} produces accurate state evolution, we eliminate the process noise and set the sensor noise standard deviation to $\sigma = 0.1$.  
	
	The simulations are performed via EPANET Matlab Toolkit~\cite{Eliades2016} on Windows 10 Enterprise with an Intel(R) Xeon(R) CPU E5-1620 v3 @3.50 GHz.  
	All codes, parameters, tested networks, and results are available on Github~\cite{wang_2020} which includes an efficient and scalable implementation of Algorithm~\ref{alg:greedy}. The details of this implementation are included in Appendix~\ref{sec:appa}.
	
	\subsection{Three-node network}\label{sec:3-node}
\begin{figure}[t]
		\centering
		\includegraphics[width=0.94\linewidth]{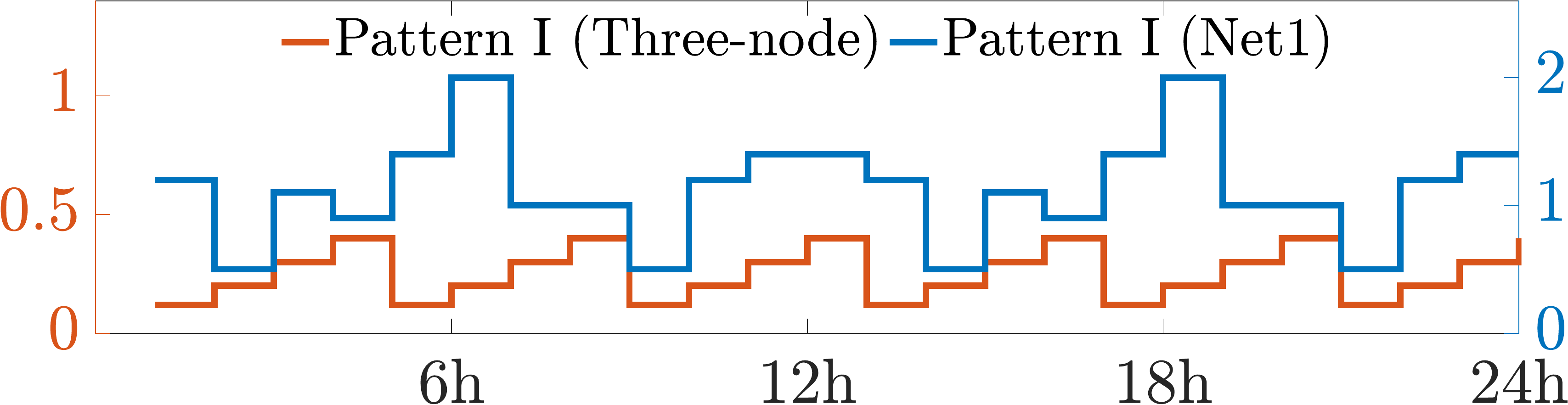}
		\caption{Pattern for Three-node and Net1 networks.}
		\label{fig:demandpattern}% THIS figure is in the LDE branch on github.
	\end{figure}

	The three-node network shown in Fig.~\ref{fig:setup}{a} includes one junction, one pipe, one pump, one tank, and one reservoir.  A chlorine source ($ c_\mathrm{R1} = 0.8$ mg/L) is installed at Reservoir 1. The initial chlorine concentrations at or in the other components are $0$ mg/L. Only Junction 2 consumes water, and its base demand is $d_{\mathrm{base}} = 2000\ \mathrm{GPM}$. The corresponding  pattern $\mathrm{Pattern\ I}$ (viewed as a row vector) for Junction 2 in 24 hours is  presented in Fig.~\ref{fig:demandpattern}.  Hence, only one demand profile for a day is computed as $\m D = d_{\mathrm{base}} \times \mathrm{Pattern\ I}$.
	The pipe is split into fixed as $s_{L_{23}} = 150$ segments, and the single hydraulic simulation interval is set to $k_f = 300 \sec$ and $T_h = 24$ hydraulic simulations are considered. To help the readers understand intuitively about the water quality modeling in state-space form and the observability (Gramian), an illustrative code including step by step comments for this small three-node network is available in our Github~\cite{wang_2020}  for the convenience of readers.

	\begin{figure}[t]
		\centering
		\subfloat[\label{fig:net1_basedemand}]{\includegraphics[keepaspectratio=true,scale=0.21]{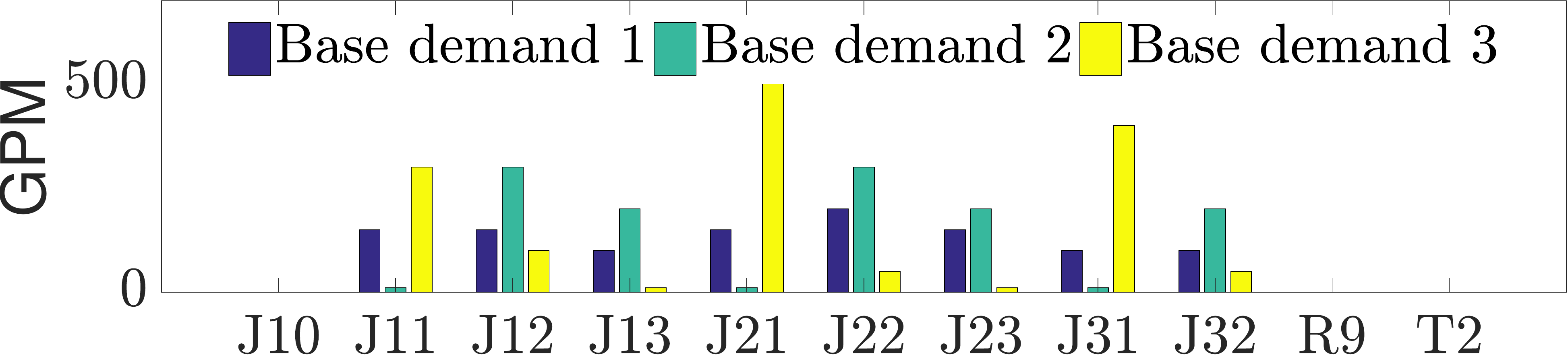}}{} 
		\subfloat[\label{fig:net1_demandpattern}]{\includegraphics[keepaspectratio=true,scale=0.20]{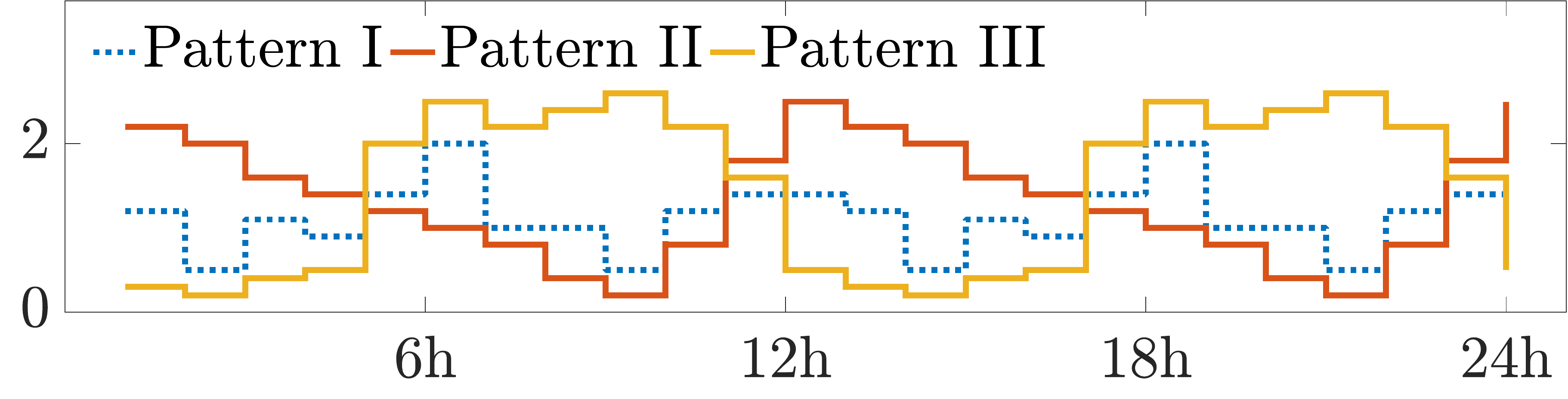}}{} 
		\caption{Different base demands (a) and demand patterns (b) for nodes in Net1.}
		\label{fig:net1_demand} 
	\end{figure}
	\begin{figure}[t]
		\centering
		\subfloat[\label{fig:SS_3_NET1_a}]{\includegraphics[keepaspectratio=true,scale=0.07]{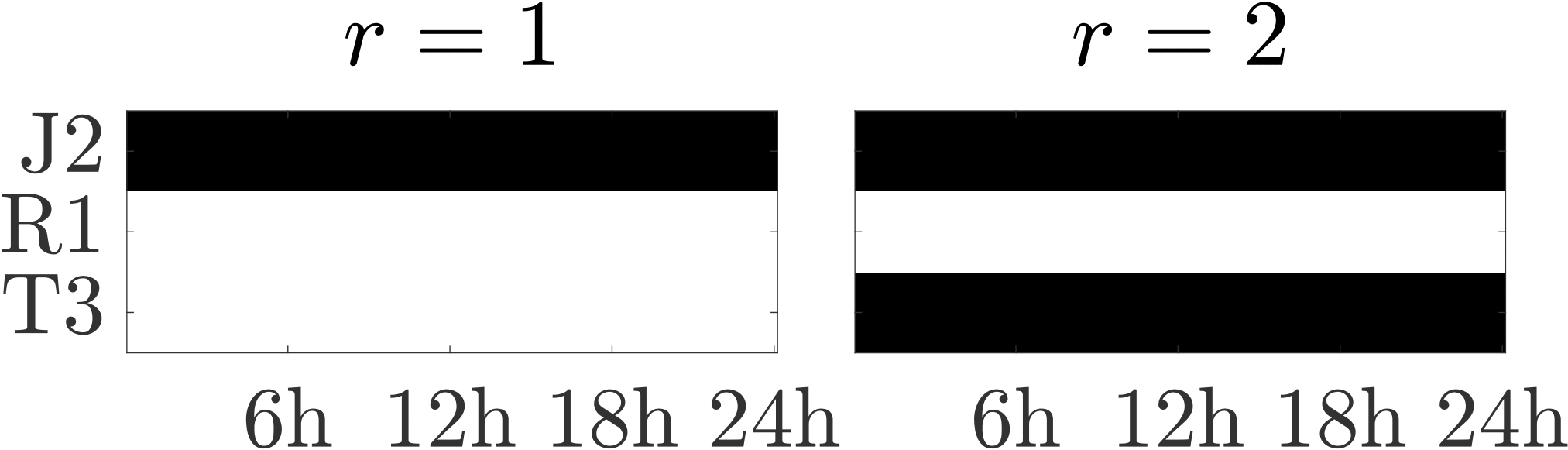}}{} 
		\subfloat[\label{fig:SS_3_NET1_b}]{\includegraphics[keepaspectratio=true,scale=0.21]{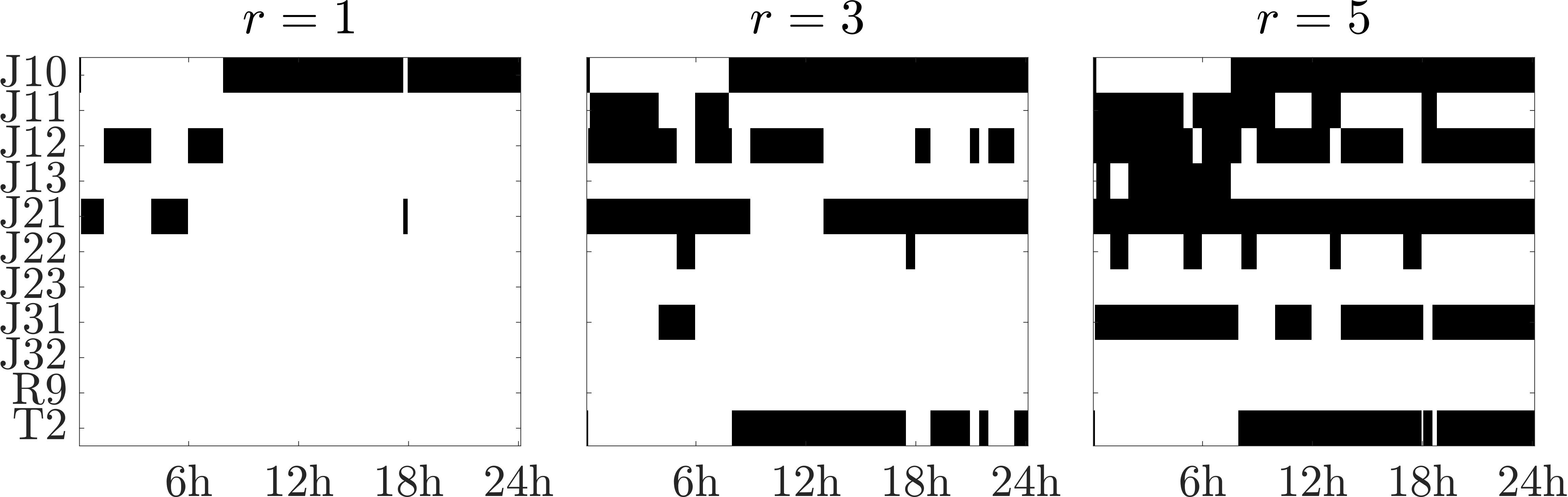}}{} 
		\caption{Sensor placement results for the three-node network (a) and Net1 (b) in 24 hours with  $k_f = 300 \sec$, $\Delta t = 5\sec$, Pattern I, Base demand $1$.}
		\label{fig:SS_3_NET1} 
	\end{figure}

	For the three-node network, there are three possible sensor locations ($\mathrm{R}1$, $\mathrm{J}2$, and  $\mathrm{T}3$); therefore, $r$ is set to $1$ or $2$ in Algorithm~\ref{alg:greedy}. The final sensor placement results are presented as Fig.~\ref{fig:SS_3_NET1_a}. When $r = 1$, $\mathrm{J}2$ is the best location or the \textit{center} of the network, and when $r = 2$, locations $\mathrm{J}2$ and $\mathrm{T}3$ are selected. To qualify the centrality or importance of a specific location during 24 hours,  \textit{occupation time} ${T}(\mc{S}) = \frac{\mathrm{Selected\ time}}{\mathrm{Total\ time}}$ is defined as a percentage of the selected time by Algorithm~\ref{alg:greedy} in a day. This measure indicates the importance of the selected sensor locations. If the sensor location does not change during 24 hours,  the occupation time would be 100\%; see Tab.~\ref{tab:sensor}. With that in mind, this 100\% figure of sensor occupation time rarely happens for any junction in larger networks---its occurrence in the three-node network is due to its simple topology. We show more interesting results with varying  occupation time in the next sections.
	
	\subsection{Looped Net1 network}\label{sec:net1}
	
	Net1 network~\cite{rossman2000epanet,shang2002particle}  shown in Fig.~\ref{fig:setup}{b}  is composed of 9 junctions, 1 reservoir, 1 tank, 12 pipes, and 1 pump.  Beyond optimal sensor placements, here we investigate the impact of the length of a single hydraulic simulation length $k_f$, L-W scheme time-step $\Delta t$, and the demand profile on the final sensor placement result. 
	This network is more complex than the three-node network because its flow direction changes and flow rates or velocities vary dramatically every hour. To balance the  performance of L-W scheme and  computational burden,  $s_{L_i}$ for each pipe is set to an integer which is the ceiling of $\frac{L_i}{ v_i(t) \Delta t}$, and dynamic number of segments setting makes $\Delta t = 5\sec$. Furthermore, If the parameter $\Delta t = 10\sec$ is needed, and this can be achieved conveniently by reducing the $s_{L_i}$ for each pipe by half.

	\subsubsection{Base case scenario and its result} The base case is considered with the following settings:  $\Delta t = 5\sec$, single hydraulic simulation $k_f = 300 \sec$, and demand profile for a single interval $\m D_k =  \mathrm{Base \  demand \ 1} \times \mathrm{Pattern\ I}$ shown in Fig.~\ref{fig:net1_demand}. There are 11 possible sensor locations (see Fig.~\ref{fig:setup}{b}), and the number of sensor locations $r$ is chosen as $[1, 3, 5]$ in~\eqref{equ:WSQP}. Similarly, we consider $24$ hours in Algorithm~\ref{alg:greedy}.  The final result is presented in Fig.~\ref{fig:SS_3_NET1_b}, and the sensor placement results in terms of occupation time $T$ are presented in Tab.~\ref{tab:sensor}.  
	
	From Fig.~\ref{fig:SS_3_NET1_b} and Tab.~\ref{tab:sensor},  when $r = 1$, $\mathrm{J}10$ in Fig.~\ref{fig:SS_3_NET1_b}  is the best sensor location most of the time ($T_{\mathrm{J}10} = 66.4\%$) and the best location switches to $\mathrm{J}12$ or $\mathrm{J}21$ occasionally ($T_{\mathrm{J}12} = 18.6\%$, $T_{\mathrm{J}21} = 14.8\%$). Hence, the solution of WQSP is $\mathcal{S}_{r=1}^* = \{\mathrm{J}10\}$ (marked as blue in Tab.~\ref{tab:sensor}). Similarly, the locations with the largest $r$ occupation time are selected as the final results when $r = 3$ and $5$. These greedy-optimal placements are given by $\mathcal{S}_{r=3}^* = \{\mathrm{J}10, \mathrm{J}12, \mathrm{J}21\}$ and  $\mathcal{S}_{r=5}^* = \mathcal{S}_{r=3}^*  \bigcup \{\mathrm{T}2,\mathrm{J}31\}$. This showcases supermodularity of the set function optimization, seeing that $\mathcal{S}_{r=3}^* \subset \mathcal{S}_{r=5}^*$.
	
		\begin{table}[t]
		\fontsize{8}{8}\selectfont
		\vspace{-0.05cm}
		\centering
		\setlength\tabcolsep{2 pt}
		\renewcommand{\arraystretch}{1.2}
		\makegapedcells
		\setcellgapes{1.0pt}
		\caption{Sensor placement results with detailed occupation time (Base case of Net1:  $\Delta t = 5\sec$, $k_f= 300\sec$ ,  Pattern I, Base   demand   1).} ~\label{tab:sensor}
		\begin{tabular}{c|c|c}
			\hline
			\textit{Network} & $r$ & \textit{Result (selected positions are in blue)} \\ \hline \hline
			\multirow{2}{*}{\makecell{\textit{Three-}\\ \textit{node}}} & $1$ & $T_{\textcolor{blue}{\mathrm{J}2}}= 100\%$ \\ \cline{2-3} 
			& $2$ &  $T_{\textcolor{blue}{\mathrm{J}2, \mathrm{T}3}}= 100\%$  \\ \hline
			\multirow{3}{*}{\makecell{\\ \textit{Net1} \\ \\ \textit{(Base case)}}} & $1$ & $T_{\textcolor{blue}{\mathrm{J}10}} = 66.4\%,$ $T_{\mathrm{J}12} = 18.6\%,$ $T_{\mathrm{J}21} = 14.8\%$ \\ \cline{2-3} 
			& $3$ & \makecell{$T_{\textcolor{blue}{\mathrm{J}10}} = 68.5\%,$ $T_{\textcolor{blue}{\mathrm{J}12}} = 56.7\%,$ $T_{\textcolor{blue}{\mathrm{J}21}} = 83.4\%$ \\ $T_{\mathrm{T}2} = 53.6\%$   } \\ \cline{2-3} 
			& $5$ & \makecell{$T_{\textcolor{blue}{\mathrm{J}10}} = 69.5\%,$ $T_{\textcolor{blue}{\mathrm{J}12}} = 87.8\%,$ $T_{\textcolor{blue}{\mathrm{J}21}} = 100\%$ \\ $T_{\textcolor{blue}{\mathrm{T}2\ }} = 65.7\%,$ $T_{\textcolor{blue}{\mathrm{J}31}} = 82.1\%,$ $T_{\mathrm{J}11} = 49.4\%$   } \\ \hline \hline
		\end{tabular}%
		%  	}
	\end{table}

	\begin{table}[t]
		\fontsize{8}{8}\selectfont
		\vspace{-0.05cm}
		\centering
		\setlength\tabcolsep{3pt}
		\renewcommand{\arraystretch}{1.2}
		\makegapedcells
		\setcellgapes{1.0pt}
		\caption{Sensor placement results considering the impacts of L-W scheme time-step $\Delta t$ and the length of the single observation time $t$ (Case A:  $\Delta t = 10\sec$, $k_f = 300\sec$; Case B: $\Delta t = 5\sec$, $k_f = 60\sec$).} ~\label{tab:sensor_impact}
		\begin{tabular}{c|c|c}
			\hline
			\textit{Network} & $r$ & \textit{Result (selected positions are in blue)} \\ \hline
			\hline
			\multirow{3}{*}{\makecell{\\ \textit{Net1} \\   \\ \textit{(Case A)} }} & $1$ & $T_{\textcolor{blue}{\mathrm{J}10}} = 71.6\%,$ $T_{\mathrm{J}12} = 12.4\%,$ $T_{\mathrm{J}21} = 14.5\%$ \\ \cline{2-3} 
			& $3$ & \makecell{$T_{\textcolor{blue}{\mathrm{J}10}} = 71.6\%,$ $T_{\mathrm{J}12} = 48.4\%,$ $T_{\textcolor{blue}{\mathrm{J}21}} = 61.2\%$ \\   $T_{\textcolor{blue}{\mathrm{T}2}} = 68.8\%$ } \\ \cline{2-3} 
			& $5$ & \makecell{$T_{\textcolor{blue}{\mathrm{J}10}} = 74.7\%,$ $T_{\textcolor{blue}{\mathrm{J}12}} = 82.7\%,$ $T_{\textcolor{blue}{\mathrm{J}21}} = 100\%$ \\ $T_{\textcolor{blue}{\mathrm{T}2\ }} = 69.2\%,$ $T_{\textcolor{blue}{\mathrm{J}31}} = 73.0\%,$ $T_{\mathrm{J}11} = 62.6\%$} \\ \hline
			\multirow{3}{*}{\makecell{\\ \textit{Net1} \\   \\ \textit{(Case B)} }} & $1$ & $T_{\textcolor{blue}{\mathrm{J}10}} = 86.5\%,$ $T_{\mathrm{J}12} = 8.3\%,$ $T_{\mathrm{J}21} = 5.1\%$ \\ \cline{2-3} 
			& $3$ & \makecell{$T_{\textcolor{blue}{\mathrm{J}10}} = 100\%,$ $T_{\mathrm{J}12} = 40.4\%,$ $T_{\textcolor{blue}{\mathrm{J}21}} = 53.6\%$ \\ $T_{\textcolor{blue}{\mathrm{T}2}} = 74.2\%$   } \\ \cline{2-3} 
			& $5$ & \makecell{$T_{\textcolor{blue}{\mathrm{J}10}} = 100\%,$ $T_{\textcolor{blue}{\mathrm{J}12}} = 85.7\%,$ $T_{\textcolor{blue}{\mathrm{J}21}} = 100\%$ \\ $T_{\textcolor{blue}{\mathrm{T}2\ }} = 78.1\%,$ $T_{\textcolor{blue}{\mathrm{J}31}} = 39.2\%,$ $T_{\mathrm{J}11} = 36.5\%$   }
			\\ \hline \hline
		\end{tabular}%
		%  	}
	\end{table}
	
	\subsubsection{The impacts of  L-W scheme time-step and the length of single observation time}
	Here, we study the impact of L-W scheme time-step and the length of single observation time parameters on the final WQSP results---in comparison with the base case from the previous section. 
	At first, only $\Delta t$ is increased from $5$ (from the base case) to $10\sec$ (Case  A). Accordingly, the number of segments of all pipes is reduced by $50\%$, while still maintaining the accuracy of LDE state-space model compared to the EPANET water quality simulation. We also define Case B by reducing $k_f$ from $300\sec$  (base case) to $60 \sec$.  The results for this experiment are shown in Tab.~\ref{tab:sensor_impact}. We observe the following: \textit{(i)} the final results are exactly the same as the ones under the base case for $r = 1,5$, and the differences are materialized only in the slightly changed occupation time; \textit{(ii)} the results under $r = 3$ are different from the base case as the solution changes from $\mathcal{S} = \{\mathrm{J}10, \mathrm{J}12, \mathrm{J}21\}$ (base case) to $\mathcal{S} = \{\mathrm{J}10, \mathrm{T}2, \mathrm{J}21\}$ (Cases A and B). This is due to the fact that the base case did not produce  a clear winner in terms of the sensor placement---the occupation times ($T_{\textcolor{blue}{\mathrm{J}12}} = 56.7\%,$ $T_{\mathrm{T}2} = 53.6\%$) are similar.   
	
	We note that even if the sensor placement strategy is changed when $r = 3$, the final performances of these three cases are comparable, and the relative error of Kalman filter performance in \eqref{equ:WSQP}  reached between Base case and Case A (Case B) is $17.2\%$ ($7.9\%$) even though the difference in $\Delta t$ is two times and the difference in $k_f$ is 5 times, which is acceptable. Hence,  one could make this preliminary conclusion: the impacts of  L-W scheme time-step $\Delta t$ and the length of hydraulic simulation $k_f$ on the final sensor placement results are negligible  assuming that the number of pipe segments (the partial differential equation space discretization parameter) is large enough to ensure the accuracy of LDE model.
	
		\begin{figure}[t]
		\centering
		\subfloat[\label{fig:net1_pattern1_base1_b}]{\includegraphics[keepaspectratio=true,scale=0.21]{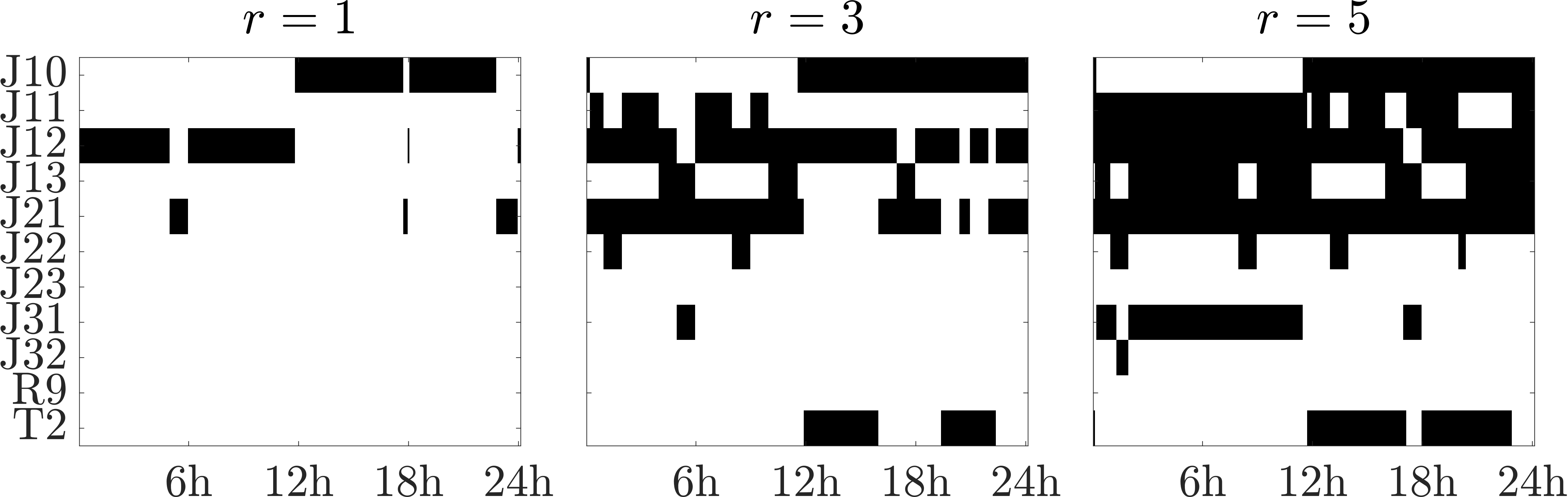}}{} 
		\subfloat[\label{fig:net1_pattern1_base1_c}]{\includegraphics[keepaspectratio=true,scale=0.21]{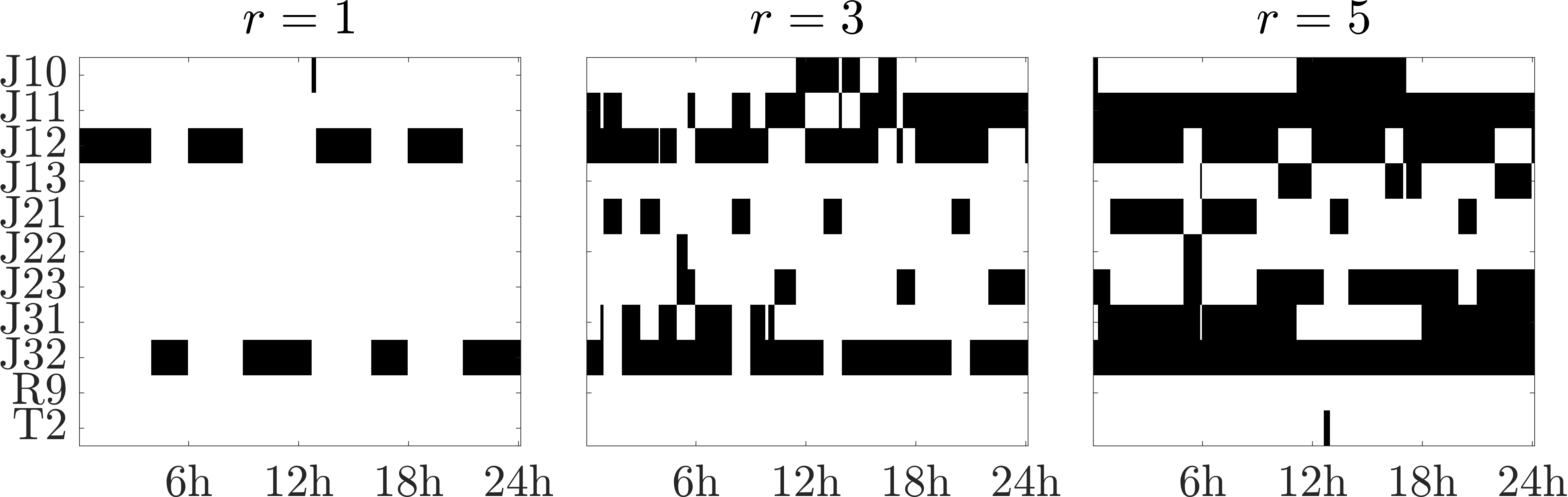}}{} 
		\caption{Sensor placement results for Net1 with  $k_f = 300\sec$, $\Delta t = 5\sec$,  Pattern I under  (a)  Base demand $2$, (b) Base demand $3$.}
		\label{fig:net1_base} 
	\end{figure}

	\subsubsection{The impact of various demand patterns}
	In this section, the impact of demand  profiles on the final sensor placement result is explored. Note that the demand in 24 hours at a node is decided by its base demand and the corresponding patterns simultaneously. Furthermore, other demand patterns could reflect other days of the weeks such as a weekend, rather than assuming a week-long demand curve.

	First,  the Pattern I is fixed as the stair-shape  in Fig.~\ref{fig:demandpattern} or the dotted line in Fig.~\ref{fig:net1_demandpattern}, and  base demands 1, 2, and 3 in Fig.~\ref{fig:net1_basedemand} are used. That is, we have $n_d = 3$ different demand profiles which is an input for Algorithm~\ref{alg:greedy}. Note that these base demands are generated for illustrative purposes.  Base demand 1 is designed to assign nearly identical base demand at each node. Base demand 2 assigns more base demands to the nodes on the right half part of the network in Fig.~\ref{fig:setup}{b}, such as $\{\mathrm{J}12, \mathrm{J}13, \mathrm{J}22,\mathrm{J}23,\mathrm{J}32\}$. Base demand 3  assigns larger base demands to the nodes on the left half part of the topology in Fig.~\ref{fig:setup}{b}, such as $\{\mathrm{J}11, \mathrm{J}21, \mathrm{J}31\}$.
	
	\begin{figure}[t]
		\centering
		\subfloat[\label{fig:net1_pattern1_b}]{\includegraphics[keepaspectratio=true,scale=0.21]{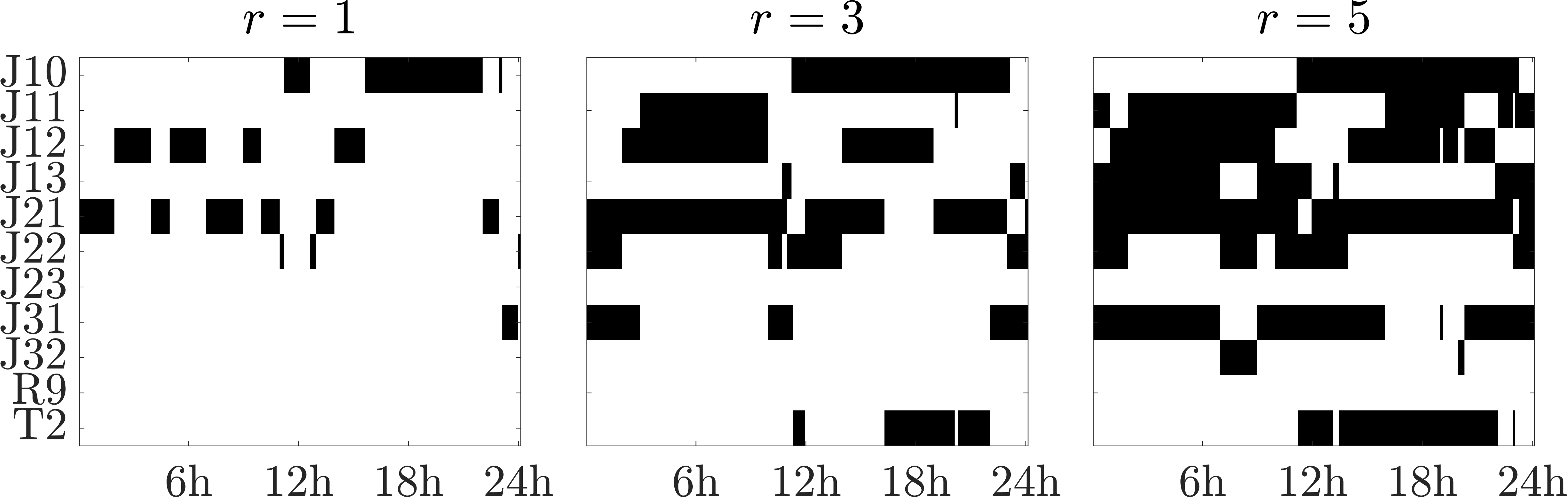}}{} 
		\subfloat[\label{fig:net1_pattern1_c}]{\includegraphics[keepaspectratio=true,scale=0.21]{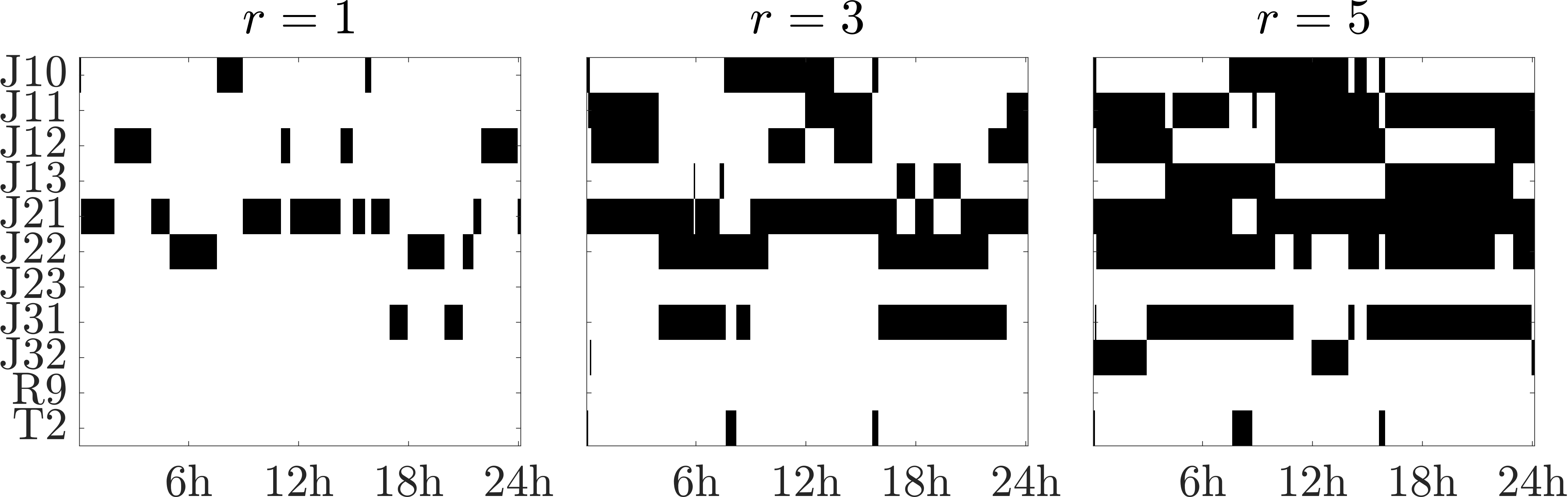}}{} 
		\caption{Sensor placement results for Net1 with  $k_f= 300\sec$, $\Delta t = 5\sec$, and Base demand $1$ under (a) Pattern II, (b) Pattern III.}
		\label{fig:net1_pattern} 
	\end{figure}

	\begin{figure}[t]
		\centering
		\includegraphics[width=0.98\linewidth]{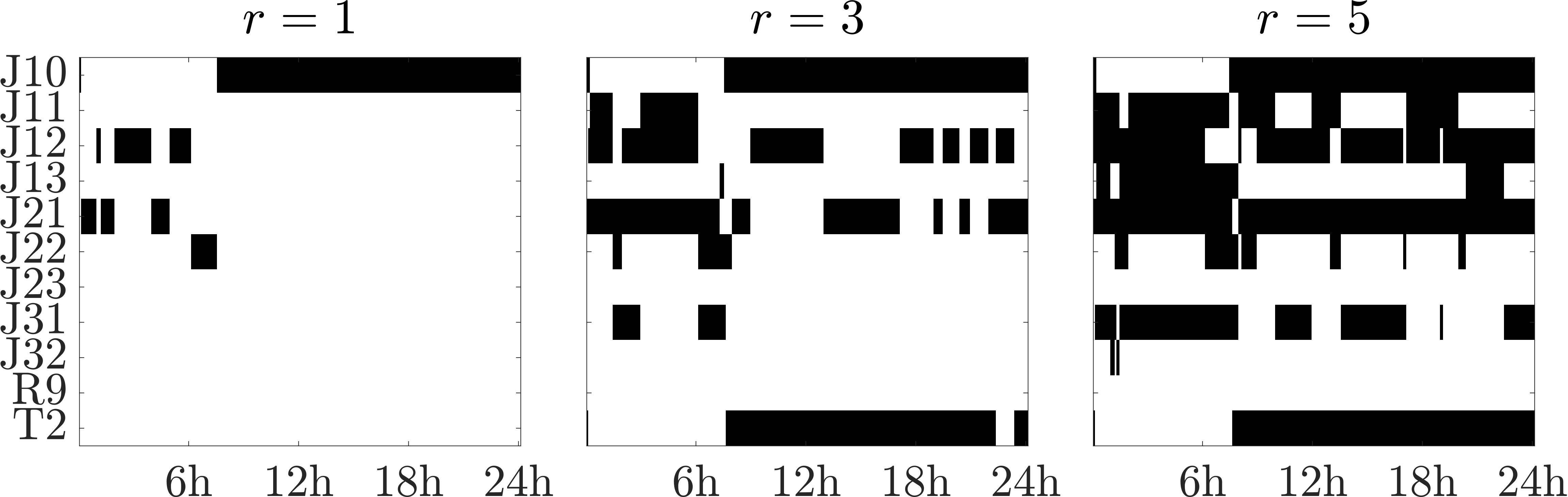}
		\caption{Final sensor placement results for Net1 with  $k_f= 300\sec$, $\Delta t = 5\sec$ consider five different demand profiles (fusion of Fig.~\ref{fig:SS_3_NET1_b}, Fig.~\ref{fig:net1_base}, and Fig.~\ref{fig:net1_pattern}).}
		\label{fig:fusion}% THIS figure is in the LDE branch on github.
	\end{figure} 
	
	The final sensor placement strategies under the three base demands 1, 2, and 3 are shown as Fig.~\ref{fig:SS_3_NET1_b}, Fig.~\ref{fig:net1_pattern1_base1_b}, and Fig.~\ref{fig:net1_pattern1_base1_c}, and the corresponding detailed occupation time are not shown for brevity. It can be observed that the greedy-optimal location switches from $\mathcal{S}_{r=1}^* = \{ \mathrm{J}10 \}$ (under base demand 1) to $\mathcal{S}_{r=1}^* = \{ \mathrm{J}12 \}$ (under base demand 3) along with changing base demand; when $r=3$, it switches from $\mathcal{S}_{r=3}^* = \{ \mathrm{J}10, \mathrm{J}12, \mathrm{J}21 \}$ (under base demand 1) to $\mathcal{S}_{r=3}^* = \{ \mathrm{J}11, \mathrm{J}12, \mathrm{J}32 \}$ (under base demand 3); when $r=5$, it switches from $\mathcal{S}_{r=5}^* = \{ \mathrm{J}10, \mathrm{J}12, \mathrm{J}21, \mathrm{J}31, \mathrm{T}2\}$ (under base demand 1) to $\mathcal{S}_{r=5}^* = \{ \mathrm{J}11, \mathrm{J}12, \mathrm{J}23, \mathrm{J}31, \mathrm{J}32 \}$ (under base demand 2). This showcases  changing base demands or different demand profiles indeed have an impact on the sensor placement, but Algorithm~\ref{alg:greedy} still returns the best placement according to the chosen metrics.

	Second, to test the impact of  patterns, Patterns I, II, and III in Fig.~\ref{fig:net1_demandpattern} are used when  base demand 1 is fixed (see Fig.~\ref{fig:net1_basedemand}). We have another $n_d = 3$ different group of demand profiles. Again, these patterns are only used for illustrative purposes to test the algorithm's performance.   It can be seen that Pattern I is relatively flatter compared with the other patterns, while Patterns II and III vary dramatically and are complementary to each other.  The final sensor placement strategies under Patterns I, II, and III are shown as Fig.~\ref{fig:SS_3_NET1_b}, Fig.~\ref{fig:net1_pattern1_b}, and Fig.~\ref{fig:net1_pattern1_c} that can also be viewed as three corresponding matrices with only zeros and ones element (not selected or selected). It can be observed that the greedy-optimal location switches from $\mathcal{S}_{r=1}^* = \{ \mathrm{J}10 \}$ to $\mathcal{S}_{r=1}^* = \{ \mathrm{J}21 \}$ and from $\mathcal{S}_{r=3}^* = \{ \mathrm{J}10, \mathrm{J}12, \mathrm{J}21 \}$ to $\mathcal{S}_{r=3}^* = \{ \mathrm{J}21, \mathrm{J}22, \mathrm{J}31 \}$. 
	With the above comparisons, we claim that both base demands and patterns would have impacts on the final sensor placement solution in Net1. In order to  quantify the similarity between two sensor placement strategies $S_1^*$ and $S_2^*$ (viewed as matrices with only zeros and ones), we define a similarity metric as $-\sum \sum \oplus (S_1^*,S_2^*)$, where $\oplus$ stands for element-wise logical operator xor. Note that this similarity metric is always a negative value, and when two matrices are the same, the largest similarity value $0$ is reached. With applying this similarity metric, Fig.~\ref{fig:net1_pattern} is closer or more similar  to Fig.~\ref{fig:SS_3_NET1_b} than Fig.~\ref{fig:net1_base}, That is, the  pattern tends  to cause less impacts than the base demand in Net1 case. This conclusion may extend to the other networks, and it is always safe to claim that varying demand profiles at each node has significant impact on the sensor placement strategy. 
	
	If we consider all discussed $n_d = 5$ demand profiles $\m D_i \in \mathbb{R}^{5 \times k_f}$ where $i = 1, \ldots, n_d$, and run Algorithm~\ref{alg:greedy}, the final sensor placement results considering Patterns I with Base demand 1,2, and 3, and Patterns II and III are shown as
	Fig.~\ref{fig:fusion}, which is the fusion of Fig.~\ref{fig:SS_3_NET1_b}, Fig.~\ref{fig:net1_base}, and Fig.~\ref{fig:net1_pattern}. The final solution $\mathcal{S}_{r=1}^* = \{ \mathrm{J}10\}$,  $\mathcal{S}_{r=3}^* = \mc{S}_{r=1}^* \bigcup \{\mathrm{J}21, \mathrm{T}2\}$, and $\mathcal{S}_{r=5}^* = \mc{S}_{r=3}^* \bigcup \{\mathrm{J}12, \mathrm{J}31\}$, thereby showcasing the greedy-optimal solution for Algorithm~\ref{alg:greedy} that exploits supermodularity of the set function optimization.

	\subsection{Net3 network}\label{sec:net3} 
\begin{figure}[t]
		\centering
		\includegraphics[width=01\linewidth]{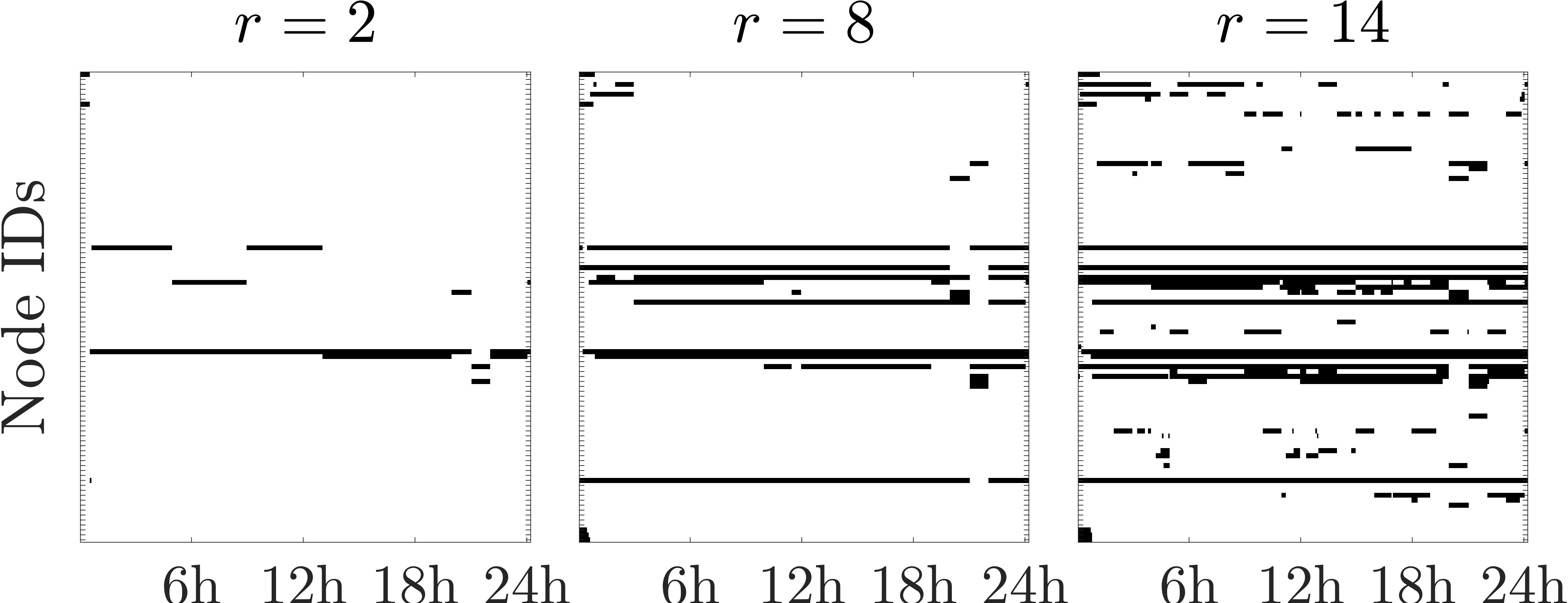}
		\caption{Sensor placement results for Net3 with $r = 2,8,14$ (95 node IDs are not shown for brevity).}
		\label{fig:net3SS}% THIS figure is in the LDE branch on github.
	\end{figure}

	In this section, the conclusions drawn  from looped Net1 network in previous section are further corroborated via the Net3 water network shown in Fig.~\ref{fig:setup}{c} with 90 junctions, 2 reservoirs, 3 tanks, 117 pipes, and 2 pumps.  The base demands of all junctions are assumed as fixed, and a relative flatten pattern (varies slightly) are tested. The results selecting  $r = 2,8,14$ from 95 node locations are shown as Fig.~\ref{fig:net3SS}, the detailed locations are presented in Tab.~\ref{tab:net3_Result}, and set $\mc{S}^*_{r=2} \subset \mc{S}^*_{r=8} \subset \mc{S}^*_{r=14}$ indicates the supermodularity property of the solution for this Net3 network. This showcases this property for even a larger network, further reaffirming the performance of the greedy algorithm. Besides that, the motivations behind testing Net3 network from a practical point of view come in two aspects, that are \textit{(i)} whether it is effective or not via adding extra sensors to reduce the Kalman filter estimation error?  and \textit{(ii)} is the strategy from Algorithm~\ref{alg:greedy}  better than random strategies?

	%, but the results are not shown due to space limitation. We observe that the occupation time of selected locations (centrality) increases, that is, the sensor placement is more fixed or time-invariant. 
	
	% Instead of setting up different patterns and comparing their results for 24 hours, a scenario which  changes the demand significantly at the 5-th and 15-th hour in a day is designed on purpose. The results selecting  $r = 2,8,14$ from 95 node locations are shown in Fig.~\ref{fig:net3SS}. From the results, the sensor placement strategy changes obviously at the 5-th and  15-th hour that verifies our conclusion thoroughly.  
	
	\begin{table}[t]
		\fontsize{8}{8}\selectfont
		\vspace{-0.05cm}
		\centering
		\setlength\tabcolsep{5 pt}
		\renewcommand{\arraystretch}{1.2}
		\makegapedcells
		\setcellgapes{1.7pt}
		\caption{Sensor placement results for Net 3.} ~\label{tab:net3_Result}
		\begin{tabular}{c|c|c}
			\hline
			\textit{Network} & $r$ & \textit{Result} \\ \hline
			\hline
			\multirow{3}{*}{\makecell{\\  \textit{Net3} \\   }} & $2$ & $\mc{S}_{r=2}^* = \{ \mathrm{J}203,\mathrm{J}204 \}$ \\ \cline{2-3} 
			& $8$ & \makecell{  $\mc{S}_{r=2}^* \bigcup \{\mathrm{J}261,\mathrm{J}163, \mathrm{J}169,\mathrm{J}173, \mathrm{J}184,\mathrm{J}206 \}$ } \\ \cline{2-3} 
			& $14$ & \makecell{ $\mc{S}_{r=8}^* \bigcup \{ \mathrm{J}208,\mathrm{J}177, \mathrm{J}179,\mathrm{J}209, \mathrm{J}20,\mathrm{J}121\}$ }
			\\ \hline \hline
		\end{tabular}%
		%  	}
	\end{table}

	\subsection{Estimation performance and comparing with random SP}
	This section computationally investigates two important issues closely related to the two motivations aforementioned: First, the performance of the state estimation and Kalman filter as the number of utilized sensors $r$ in the water network varies. The second issue is whether a uniform (i.e., placing a sensor every other junction) or random sensor placement strategy yields a comparable performance---in terms of the state estimation metric---when comparison with the greedy-optimal presented in Algorithm~\ref{alg:greedy}. Both issues are investigated for the larger network Net3. 
	
\begin{figure}[t]
		\centering
		\subfloat[\label{fig:net3Reached}]{\includegraphics[keepaspectratio=true,scale=0.33]{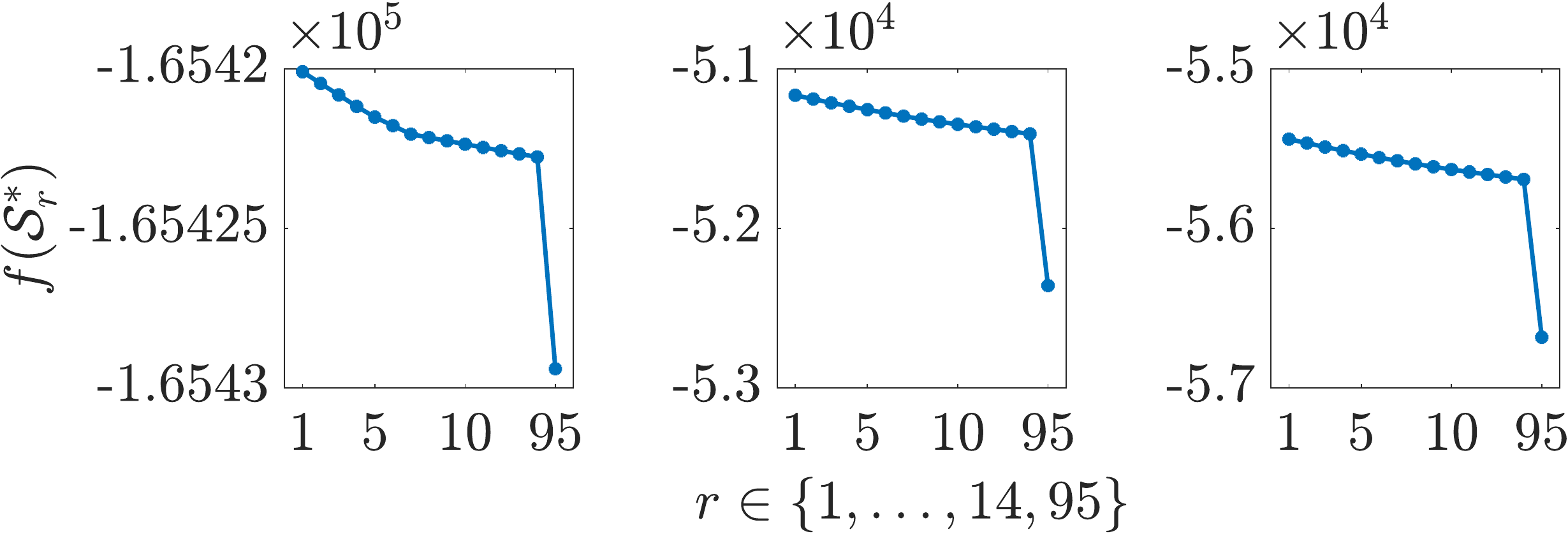}}{} 
		\subfloat[\label{fig:net3Random}]{\includegraphics[keepaspectratio=true,scale=0.23]{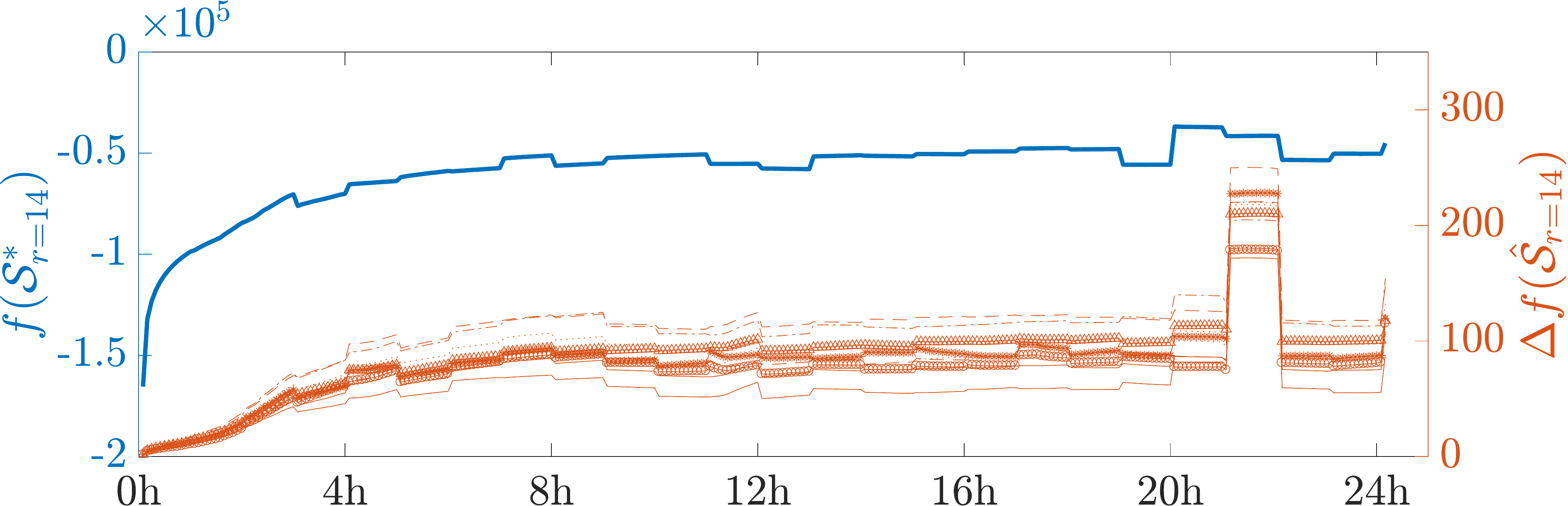}}{} 
		\caption{Kalman filter performance $f(\mc{S}_{r}^*)$ with $r = \{1,\ldots,14,95\}$ when $T_h = 0^\mathrm{th}, 10^\mathrm{th}, 20^\mathrm{th}$ hour (a), performance $f(\mc{S}^*_{r=14})$ for 24 hours (blue line in (b)), and the relative performance of ten randomized sensor placements $\Delta f(\hat{\mc{S}}_{r=14})$  (red lines in (b)).}
		\label{fig:net3_performance} 
	\end{figure}

	First, the relationship between performance of the Kalman filter $f(\mc{S}_r)$~\eqref{equ:obsmetric} and the number of sensors is shown as Fig.~\ref{fig:net3Reached}. Interestingly, Kalman filter performance $f(\mc{S}_r)$ decreases roughly linearly as the number of sensors $r$ increases from $1$ to $14$ for three different hydraulic simulations. Specifically, Fig.~\ref{fig:net3Reached} showcases the performance of the greedy-optimal solution when $r$ is fixed in Algorithm~\ref{alg:greedy} with fixed hydraulic profiles ($T_h = 0^\mathrm{th}, 10^\mathrm{th}, 20^\mathrm{th}$ hour) i.e., the three figures in Fig.~\ref{fig:net3Reached}  show similar trend for three different hydraulic profiles. The best performance or lower bounds under the corresponding cases are reached when all sensor locations are selected ($r = 95$). This indicates that one would not expect a large improvement of Kalman filter performance via increasing the number of sensors even the locations of added sensors are all greedy-optimal.

	Furthermore, the time-varying Kalman filter performance $f(\mc{S}^*_{r = 14})$ for 24 hours is depicted via the blue line in Fig.~\ref{fig:net3Random}.  the  performance value can easily reach $10^{5}$ level for this relatively large-scale network due to \textit{(i)} the large dimension of $\m z$ ($n_z = 3.066 \times 10^6$), \textit{(ii)}  covariance matrix $\mathbb{C}$ with tiny diagonal element (i.e., $5 \times 10^{-3}$), and \textit{(iii)} the typical value of $k_f$ is 200 in Net 3 resulting in $\m W_o$ with huge value element in \eqref{equ:closedform}. Moreover, the trend of the blue line is decided by the hydraulic profile such as the flow rates for 24 hours, the plot of flow rates are not shown for brevity.
	
	To address the second issue, we showcase the performance of a random sensor placement with a fixed number of sensors $r=14$.
	%The  performance value can easily reach $-10^{5}$ due to \textit{(i)} high dimension $n_z$ (i.e., $15,430$), \textit{(ii)}  covariance matrix $\mathbb{C}$ with tiny diagonal element (i.e., $5 \times 10^{-3}$), and \textit{(iii)} $\m W_o$ with huge value element in \eqref{equ:closedform}. 
	Specifically, ten random sensor placements are generated for every hydraulic simulation. To quantify the performance of the proposed optimal placement, we define the relative performance of a random placement strategy $\hat{\mc{S}}$ as $\Delta f(\hat{\mc{S}}_{r=14}) = f(\hat{\mc{S}}_{r=14}) - f(\mc{S}^*_{r=14})$. A smaller value of $\Delta f(\hat{\mc{S}}_{r=14})$ implies a  better optimal placement. The red lines in Fig.~\ref{fig:net3Random} are the relative performance of ten different randomizations---all of them are greater than zero showcasing a much better state estimation performance through the greedy algorithm. Even though the differences of performance are only 100-200 on average, the actual Kalman filter performance is orders of magnitude better due to fact that the $\log \det$ function is used to quantify the state estimation accuracy. That is, the  $\mc{S}^*_{r=14}$ obtained from Algorithm~\ref{alg:greedy} performs significantly better than any random strategy $\hat{\mc{S}}^*_{r=14}$.

	\section{Conclusions, Paper Limitations, and Future Directions}
	
	The paper presents a new computational method that results in sensor placements of WQ sensing devices in water networks. The method exclusively focuses on the WDN observability in regards to the WQ dynamics.  After thoroughly testing three networks, we  summarize the findings. 
	First, the impacts of choosing L-W scheme time-step $\Delta t$ (or the number of segments $s_L$) and the length of a single hydraulic simulation $k_f$ on the sensor placement strategy is minor and can be neglected. Second, our proposed method can be applied to practical networks to determine sensor placement strategy because in practice historical data for demand patterns are available, thereby furnishing the sensor placement with the most common demand patterns. Hence, there is a possibility that  the optimal sensor placement in terms of occupation time obtained would relatively be \textit{time-invariant}. Third, the algorithm verifies the supermodular nature of the advocated set function optimization as corroborated via different test cases on three different networks. Fourth, and even if demand patterns change significantly, the algorithm can still be used to obtain a sensor placement that optimizes the sensor occupation time. 
	
	The paper \textit{does not} advocate for \textit{only} looking at state estimation metrics for water quality sensor placement. As mentioned in Section~\ref{sec:literature}, a plethora of social and engineering objectives are typically considered in the literature to solve the WQSP. To that end, it is imperative that the proposed approach in this paper be studied in light of the other metrics and objectives discussed in the literature (such as minimizing the expected population and amount of contaminated water affected by an intrusion event). Consequently, an investigation of balancing engineering and state estimation objectives with more social-driven ones is needed. Hence, the objective of this paper is to provide a tool for the system operator that quantifies network observability vis-a-vis water quality sensor placements.  The water system operator can also balance the objective of obtaining network-wide observability with these other metrics. Future work will focus on this limitation of the present work, in addition to considering multi-species dynamics that are nonlinear in the process model, which necessitate alternate approaches to quantify observability of nonlinear dynamic networks. This will also allow examining the potential reaction between contaminants and chlorine residuals that the sensors are monitoring.
	
		\section*{Data Availability Statement}
	
	Some or all data, models, or code used during the study were provided by a third party. Specifically, we provide a Github link that includes all the models, the data, and the results from the case study~\cite{wang_2020}.
	
	\section*{Acknowledgment}
	
	This material is based upon work supported by the National Science Foundation under Grants 1728629, 1728605, 2015671, and 2015603. 
	\appendix
	
	\section{Scalability and Efficient Algorithm Implementation}~\label{sec:appa}
	
	This section presents a brief discussion on an efficient implementation of Algorithm~\ref{alg:greedy} in light of the large-scale nature of the problem. This nature is due to the size of water networks, but mainly due to the space discretization of the PDE. This results in a large state-space dimension for the LDE model~\eqref{equ:ltv}. Considering Net3 for example with a single hydraulic simulation (i.e., $t = 5$ minutes or equally $k_f = 300$ time-steps when  $\Delta t$ = $1$ second to reach a decent performance in L-W scheme),  the number of segments of different pipes is set as $s_L = 1000$. Given these figures, the corresponding dimension of the state-space model~\eqref{equ:de-abstract1} and~\eqref{equ:ltv} is $n_x = 117,099$. That is, $\m A \in \mathbb{R}^{117,099 \times 117,099 }$, $\m C \in \mathbb{R}^{95 \times 117,099 }$ in~\eqref{equ:ltv}, $\mathcal{\m O}(k_f) \in \mathbb{R}^{95 k_f \times 117,099 k_f}$ in~\eqref{equ:ymeasurement}, and $\m W_o \in \mathbb{R}^{117,099 k_f \times 117,099 k_f}$ in~\eqref{equ:closedform} with $k_f = 300$. Next, we discuss the balance between the accuracy of our LDE and the computational burden. 
	
	From the above example, the problem dimension is determined via number of the water quality simulation time-steps $k_f$, the number of pipes $n_{\mathrm{P}} $, and the number of segments of each single pipe $s_{L}$.  Note that  $k_f = t/\Delta t$, where $\Delta t \leq \min(\frac{L_i}{s_{L_i} v_{i}(t)})$ for all  $i \in \mathcal{P}$. Ideally, parameter $s_{L_i}$ should be as large as possible to ensure accuracy of LDE model. After fixing the length of the interval $t$,  parameter $k_f$ ($\Delta t$) should be as small (large) as possible to reduce the dimension of $\mathcal{\m O}$ in~\eqref{equ:ymeasurement}. This subsequently reduces the computational burden of finding the  $\log$ $\operatorname{det}$ of a large-scale matrix in~\eqref{equ:closedform}---and hence improves the computational tractability of Algorithm~\ref{alg:greedy}. Evidently, there exists a conflict between increasing of the accuracy of the LDE model (via increasing $s_L$ for each pipe) and reducing computational burden (via decreasing $k_f$, increasing $\Delta t$ or, equally, decreasing $s_L$) in a single hydraulic simulation.   
	We next show simple approaches which can alleviate this conflict, significantly reduce the computational burden while maintaining the accuracy, and yield a scale algorithmic implementation.
	
	\begin{itemize}
		\item First, the state-space matrices $\m A$ and $\m C$, which are the major component of the dynamic water quality model and all subsequent matrices and Algorithm~\ref{alg:greedy}, are extremely sparse. In fact, more than 99.9\% of the entries of these matrices are zeros. Thus, matrices $\m A$ and $\m C$  can be expressed in the sparse matrix form thereby reducing the computational burden and the needed memory by many orders of magnitude. We use this in our Github codes.
		\item Second, to reduce the number of time-steps $k_f$ or increase $\Delta t$, the dynamic number of segments of each pipe should be adopted due to the fact that the $\Delta t$ is related with  pipe length $L_i$ and its velocity $v_{i}(t)$ which depends on the hydraulic simulation and demand profile. That is, for the short (long) Pipe $i$ with large (small) velocity for interval $t$, the number of segments $s_{L_i}$ can be chosen as a relatively small (large) value that is still enough to ensure the accuracy of L-W scheme. With varying number of segments, the dimension of $\m A$ (i.e., $n_x$) varies in different hydraulic simulations. The interested reader is referred to our Github~\cite{wang_2020} for the details of this implementation.  For example, if the velocities during interval $t_1$ are two times less than the ones during interval $t_2$, then the size of $n_x$ can be reduced by half in $t_2$.  
		\item Third, the WQSP problem parameters~\eqref{equ:WSQP}  in each single hydraulic simulation are independent on each other.  This is due to $\m A$ and $\m C$ in~\eqref{equ:ymeasurement} are time-invariant in a single interval. After obtaining initial conditions and matrices $\m A$, $\m C$ offline, Algorithm~\ref{alg:greedy} can be implemented through parallel computing.  That is, multi-intervals can be calculated simultaneously on a multi-core computer. Moreover, the bottleneck of Algorithm~\ref{alg:greedy} is located in calculating the $\log \operatorname{det}$ of large-scale matrix~\eqref{equ:closedform}.  To that end, we adopt the LU and Cholesky decompositions~\cite{trefethen1997numerical}  to accelerate the evaluation of the  $\log \operatorname{det}$ objective function for different sensor placements.
		\item Fourth, it is clear that reducing the length of a single hydraulic simulation $t$ would be another effective trick, since it would reduce the dimension of $\mathcal{\m O}(k)$ and $\m W_{O}(k)$ significantly according to~\eqref{equ:closedform}.
	\end{itemize}
	Finally, it is worth mentioning that Algorithm~\ref{alg:greedy} is indeed offline seeing that it solves a \textit{placement} problem of water quality sensors: sensors that cannot have varying geographic locations. Powerful computational resources, at the disposal of water network operators, can hence be used to run the algorithm. 
	
	We present the tested computational time for running a horizon of 24 hours for the tested three networks. For the three-node network, it takes $45.7 \sec$ to terminate Algorithm~\ref{alg:greedy}; for Net1 with larger discretization time-step $\Delta t = 10 \sec$, the computational time is $129.6 \sec$, and the computational time becomes $188.1 \sec$ when the time-step is reduced by half, that is, $\Delta t = 5 \sec$. When we change a new demand profile for the Net1 network, the computational time can reach $555.6 \sec$; for the Net 3 network it takes almost four hours for the entire 24 hour simulation horizon.  The reasons of the varying computational time are: (i) special demand profiles may result in time-consuming LU or Cholesky decompositions, (ii) the discretization time-step $\Delta t$ and length of single observation time $k_f$ have an impact on the size of $\m W_o$ and result in different computational time, (iii) and the tested computer have different numbers of cores and capacities of memory.

	\bibliographystyle{IEEEtran}
	\bibliography{bibfile2.bib}
\end{document}

%% file: preambleT.tex
\usepackage{epsfig,color,amsmath}
 
\usepackage{wrapfig}
\usepackage{setspace}  
\usepackage{xcolor}
\usepackage{epstopdf}
\usepackage{amssymb}
\usepackage{url}
\usepackage{mathtools}
\usepackage{enumitem}
\usepackage{multirow}
\usepackage{makecell}
\usepackage{amsmath}
\usepackage{stackrel}
\usepackage{booktabs}
\usepackage[flushleft]{threeparttable}
\usepackage{makecell}

\usepackage{subfig}
\usepackage{graphicx}

\usepackage[linesnumbered,lined,boxed,commentsnumbered,ruled,longend]{algorithm2e}


\setlist[itemize]{leftmargin=*}

\DeclareMathOperator{\rank}{rank}

\DeclareMathOperator*{\minimize}{minimize}

\DeclareMathOperator{\subjectto}{subject\ to}

\makeatother

\newcommand{\m}{\boldsymbol}
\allowdisplaybreaks[4]
\usepackage[colorlinks=true,citecolor=blue,linkcolor=black,urlcolor  = blue,anchorcolor = blue]{hyperref}

\newcommand{\mc}[1]{\mathcal{#1}}
\newcommand{\mbb}[1]{\mathbb{#1}}

\usepackage{amsthm}

\usepackage[usestackEOL]{stackengine}
\stackMath
\usepackage[framemethod=TikZ]{mdframed}
\usepackage{bm}
\mdfdefinestyle{MyFrame}{%
	linecolor=black,
	outerlinewidth=1.5pt,
	roundcorner=1.5pt,
	innerrightmargin=5pt,
	innerleftmargin=5pt,}

\usepackage{tikz}
\usetikzlibrary{matrix,positioning,decorations.pathreplacing}

\usepackage[export]{adjustbox}